\documentclass[journal]{IEEEtran}
\ifCLASSINFOpdf
   \usepackage[pdftex]{graphicx}
\else
\fi
\usepackage{amsmath}

\usepackage{hyperref}

\usepackage{cite}

\usepackage{multirow}
\usepackage{multicol}
\usepackage{mathtools}

\usepackage{amsmath,epsfig}
\usepackage{amsfonts}
\usepackage{url}
\usepackage{footnote}

\newcommand{\ty}{\tilde{y}}
\newcommand{\tx}{\tilde{x}}
\newcommand{\tn}{\tilde{n}}

\begin{document}
%
\title{Deep Learning Methods For Synthetic Aperture Radar Image Despeckling: An Overview Of Trends And Perspectives}
%
%
%

\author{Giulia Fracastoro, Enrico Magli, Giovanni Poggi, Giuseppe Scarpa, Diego Valsesia, Luisa Verdoliva}

\maketitle

\begin{abstract}
Synthetic aperture radar (SAR) images are affected by a spatially-correlated and signal-dependent noise called ``speckle'', which is very severe and may hinder image exploitation. Despeckling is an important task that aims at removing such noise, so as to improve the accuracy of all downstream image processing tasks. The first despeckling methods date back to the 1970's, and several model-based algorithms have been developed in the subsequent years. The field has  received growing attention, sparkled by the availability of powerful deep learning models that have yielded excellent performance for inverse problems in image processing. This paper surveys the literature on deep learning methods applied to SAR despeckling, covering both the supervised and the more recent self-supervised approaches. We provide a critical analysis of existing methods with the objective to recognize the most promising research lines, to identify the factors that have limited the success of deep models, and to propose ways forward in an attempt to fully exploit the potential of deep learning for SAR despeckling. 
\end{abstract}

\section{Introduction}

Synthetic Aperture Radar (SAR) data are generated through an active and coherent sensing process, whereby radar echo returns of ground targets are acquired (the so-called raw data) and later focused into a fully developed image. 
SAR images represent an important complementary source of information with respect to optical images. While the latter are more easily interpretable and gather data in many spectral bands, allowing e.g. for a detailed classification of ground covers, SAR images provide information on the geometry of the observed scene allowing also, in interferometric and polarimetric modalities, the accurate retrieval of topography and 3D structures. Moreover, SAR image acquisition does not depend on sunlight or climatic conditions, ensuring a continuous all-day all-weather coverage.

\begin{figure*}
    \centering
    \includegraphics[width=0.36\textwidth]{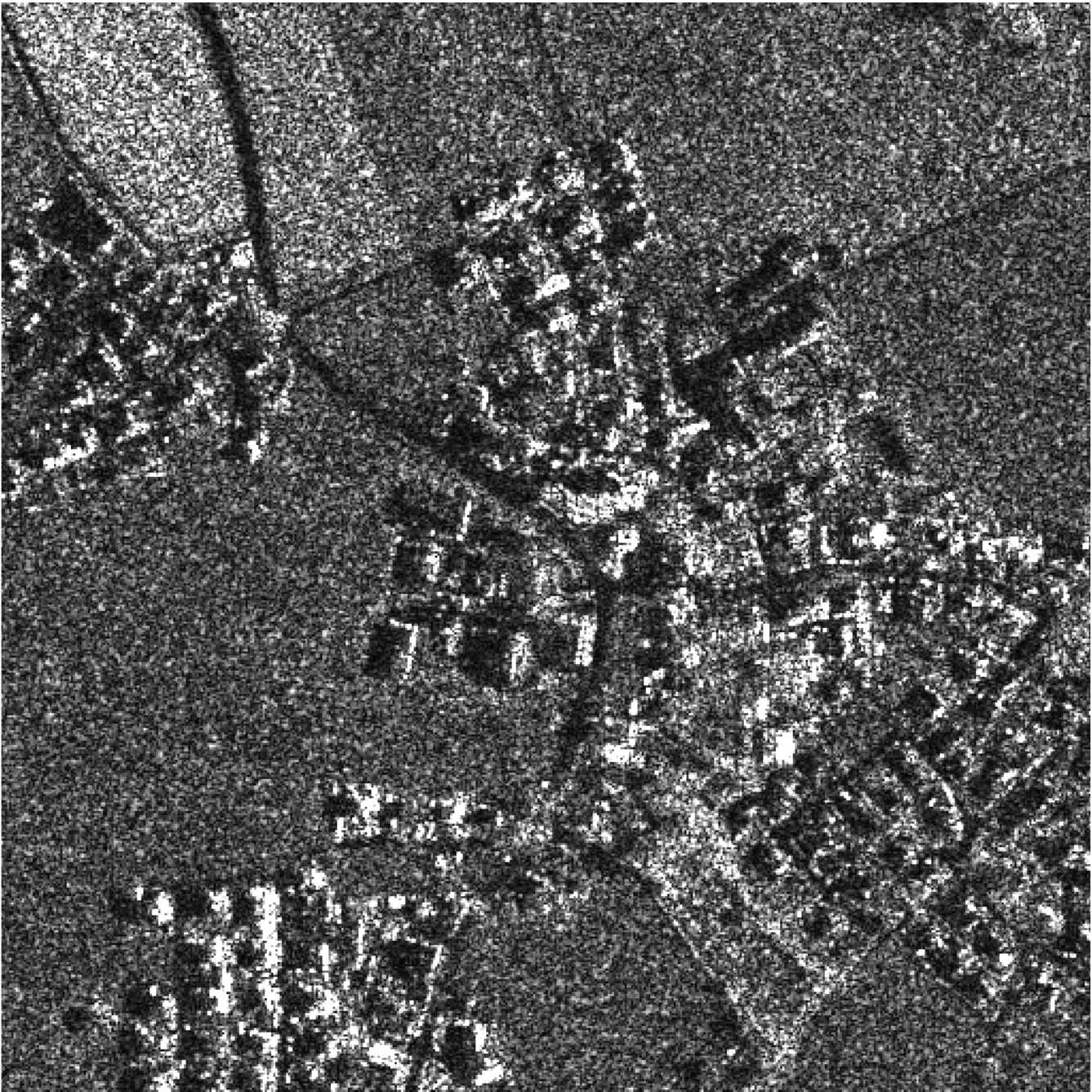}
    \includegraphics[width=0.36\textwidth]{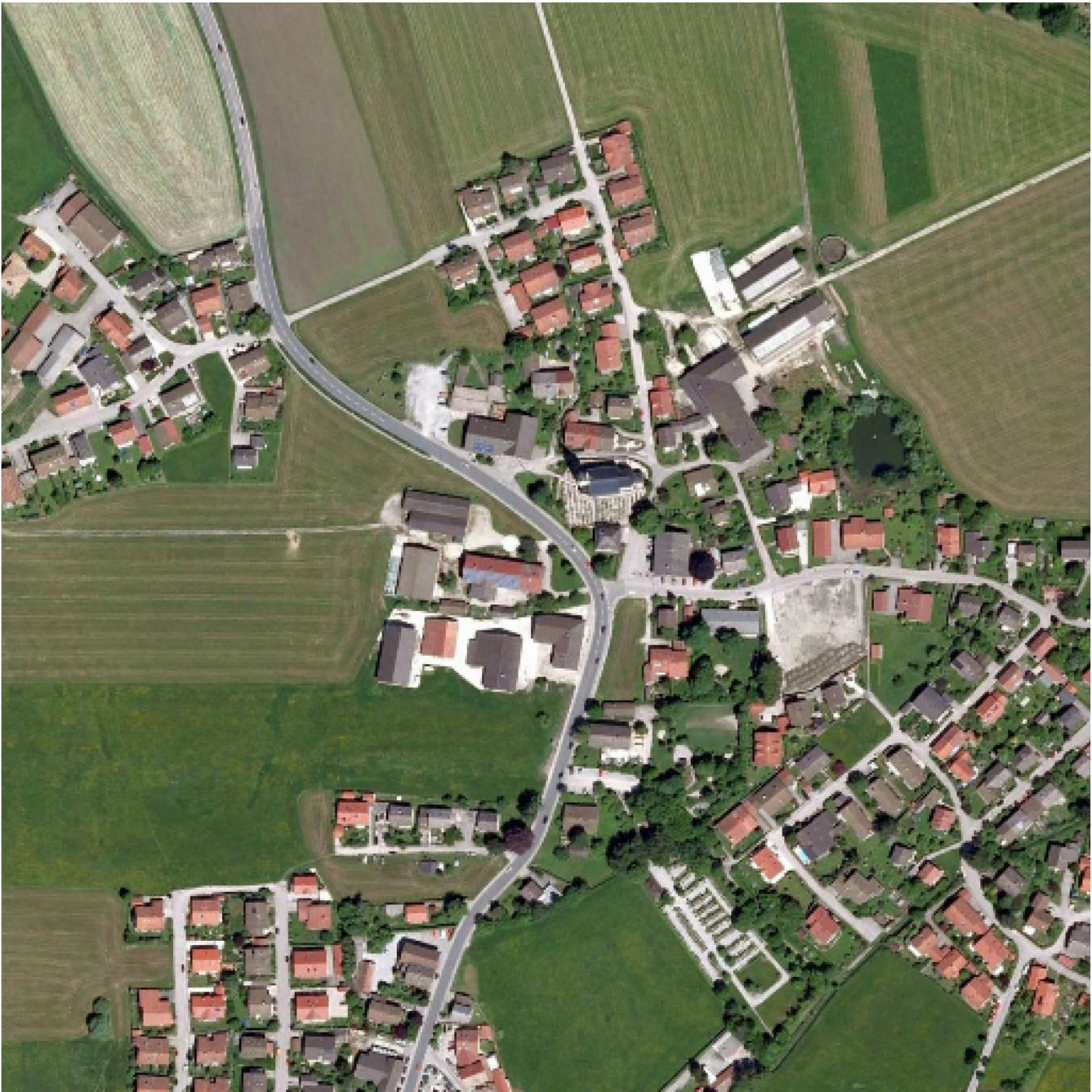}
    \caption{Left: a single-look TerraSAR-X image acquired over Rosenheim (D).
    Right: an optical image of Rosenheim downloaded from Google-Earth Pro, co-registered with the SAR master.
    The noise intensity is extremely large for the SAR image, virtually zero for the optical image.}
    \label{fig:speckle}
\end{figure*}

However, as in other coherent imaging systems such as medical ultrasound and tomography, the coherent sensing process entails that the intensity of SAR images is affected by a particular type of noise called ``speckle''. Unlike the Gaussian noise typically affecting optical images, speckle noise is spatially correlated and signal-dependent, and appears as a grainy texture superimposed to the image (see e.g. Fig.\ref{fig:speckle}), which greatly affects its interpretability and scientific exploitation. Therefore, improving SAR image quality by reducing noise has a significant positive impact on all downstream tasks employing such images, from visual inspection to environmental monitoring, surveillance, detection of anomalies and hazards, just to mention a few. 

The interest for speckle rejection has been intense since the early days, as testified by papers on multi-looking \cite{porcello1976} and Wiener filtering \cite{kondo1977} dating back to the 1970s.
Several effective methods \cite{lee1980spatialadaptive,kuan1985spatialadaptive,frost1982spatialadaptive,lopes1990spatialadaptive} were proposed already in the 1980s, based on spatial-adaptive filtering.
The key idea is to adapt the estimation window based on local image statistics, so as to deal with signal non-stationarity.
Even so, speckle reduction is accompanied by a clear loss of resolution.
A further class of methods \cite{xie2002wavelet, argenti2002wavelet, solbo2004wavelet, bianchi2008wavelet} appeared in the 2000s with the advent of the wavelet transform.
By leveraging the good separation of signal and noise in the wavelet domain, they combine wavelet shrinkage with models and tools typical of image processing.
Despite the improved speckle rejection, visible filtering artifacts, such as ringing, are frequent.
Variational methods \cite{aubert2008variational,bioucasdias2010variational,shi2008variational}, instead,
formulate image despeckling as an optimization problem, looking for the latent image which best explains the observed noisy image.
To remove speckle, an objective function is minimized which compounds a data-fitting term with a regularization term, typically based on total variation.
Again, significant artifacts are observed, like signal oversmoothing and staircasing.
Nonlocal methods \cite{deledalle2009nonlocal, Parrilli2012SARBM3D, Cozzolino2014FANS, deledalle2015nonlocal} represent arguably the current state of the art.
The core idea is to select the best predictors of the target pixel through a suitable measure of similarity based on the local context \cite{deledalle2014exploiting}.
By combining the nonlocal approach with other image processing tools, a good speckle rejection is obtained with limited (though not absent) artifacts and loss of resolution.

The advent of deep learning has revolutionized the approach to many image processing tasks. After the stunning results obtained by the AlexNet convolutional neural network (CNN) in the ImageNet classification challenge in 2012 \cite{alexnet2012}, a flurry of methods have appeared, attempting to tackle various image processing problems, including image restoration,  using deep learning methods. In most cases, also owing to the ample availability of training images, such deep learning methods have been able to learn increasingly sophisticated image models, thereby achieving significantly better performance than previous methods. For example, for optical image denoising, state-of-the-art data-driven methods such as DnCNN \cite{zhang2017beyond}, and more recently non-local neural networks \cite{plotz2018neural,liu2018non,valsesia2020deep}, have outperformed by a large margin model-based approaches such as BM3D \cite{dabov2007image} and its variants.

These promising results have encouraged researchers to address the SAR despeckling problem using deep learning models. While such models have indeed yielded performance improvements, the obtained gains have fallen somewhat short of the expectations raised by the huge success on optical images. Why is this the case, and how can we improve further?

This paper provides a survey of deep learning methods for SAR image despeckling with the objectives to assess the state of the art in this field, to evaluate what has and has not worked so far, to identify the factors that have limited the success of deep learning-based approaches, and to propose ways forward in an attempt to fully exploit the potential of deep learning for SAR despeckling. Underpinning this analysis is the notion that speckle is a rather complex and difficult to handle  type of noise; direct application of methods developed for additive Gaussian noise and optical images, e.g. vanilla CNNs, are not going to work well. We organize the existing body of work according to the neural network architectures, loss functions, training and testing strategies, and training data, presenting the most important works and identifying their key contributions, and eventually outlining research perspectives.

The paper is organized as follows. Two sections provide an introduction to the speckle noise (Sec. \ref{sec:primer_speckle}) and to deep learning (Sec.  \ref{sec:primer_deep}) respectively, providing background material and notation. Sec. \ref{sec:SAR_data} introduces SAR data. First, the problem of preparing a good dataset for despeckling is addressed, outlining several possible strategies from simple simulation to multitemporal fusion. Then, available sources of SAR data are presented and discussed, along with the related problems. Next, the body of the paper is concerned with the critical analysis of existing despeckling methods. First, supervised methods are addressed. These are classified according to whether they are directly employed as denoising models or they are learnable components of model-based techniques. The analysis of the former category further explores various types of architecture (Sec. \ref{sec:supervised_architectures}), with specific focus on how the architecture can reflect different noise models. 
Then, training and testing strategies are discussed in Sec. \ref{sec:supervised_training} with emphasis on the limitations of supervised learning. Having completed the presentation of supervised methods, Sec. \ref{sec:selfsupervised} deals with the recent approach of self-supervised training, explaining how this approach does away with the need of {\em clean} images during the training process, opening the way to a fuller exploitation of available SAR data. Finally, Sec. \ref{sec:future_directions} wraps up the paper by discussing the main limitations of existing approaches, and outlining possible avenues of future research that may better harness the representation power of deep models for SAR image despeckling.

\section{Primer on SAR speckle} \label{sec:primer_speckle}

A synthetic aperture radar is an active coherent imaging system.
It illuminates the scene at a given wavelength and, by means of a sophisticated focusing process, collects the returns back-scattered by the surface.
The amplitude of the signal associated with the individual resolution cell depends not only on the reflectivity of the illuminated surface but also on its geometry and its roughness at the scale of the wavelength.
Among the infinite possible combinations of these factors \cite{Frery1997amodel}, two extreme cases are worth emphasizing, fully developed speckle and specular reflectors.

In the first case, the cell includes a large number of independent elementary scatterers.
Therefore, the received signal is obtained as a sum of many contributions,
with amplitudes which are all proportional to reflectivity, but phases which, due to the different optical paths, can be regarded as random variables (RVs) uniform in $[0,2\pi]$.
By the central limit theorem,
the in-phase and quadrature components of the received signal are independent identically distributed Gaussian RVs, with zero mean and a variance that depends on the reflectivity.
Eventually, the intensity (square of the amplitude) of the received signal can be written according to a {\em multiplicative noise model}
\begin{equation}
    y= xn
    \label{eq:multiplicative}
\end{equation}
The signal component, $x$, accounts for the reflectivity of the material,
while $n$ is a random variable with unit-mean exponential probability density function (pdf)
\begin{equation}
    p(n) = \exp(-n) {\rm u}(n)
    \label{eq:exponential}
\end{equation}
with ${\rm u}(\cdot)$ the unit step function,
which accounts for the random aggregation of the elementary contributions, and is commonly regarded as noise.
This is the well-known Goodman's stochastic model for fully developed speckle.
It fits a large variety of situations, especially involving natural land covers with rough surfaces, which do contribute a great many elementary scatterers in a single resolution cell.

The other extreme case, instead, occurs often in urban areas or, more in general, when man-made objects are present in the cell.
In these circumstances, smooth perpendicular surfaces are often present, forming orthogonal dihedral or even trihedral angles, which act as specular reflectors.
Then, the intensity of the signal becomes extremely large, and is not affected by noise.
Obviously, Goodman's model does not hold in this case.

In Fig.\ref{fig:speckle} we show, on the left, an example SAR image where both cases occur:
agricultural fields, characterized by large areas with homogeneous reflectivity and fully developed speckle,
and a dense residential area, with a number of buildings, characterized by intense, noiseless, specular reflection as well as shadow regions.
 In the same figure, we also show, on the right, an image of the same scene acquired by an optical sensor, co-registered with the SAR image.
The comparison makes clear that the SAR image is much noisier than the optical one.
In homogeneous areas, in particular, the SAR image has a salt-and-pepper appearance, which justifies the name speckle noise\footnote{Strictly speaking,
speckle is not noise, since it conveys itself useful information on the roughness of the imaged surface.
However, for most applications this property is immaterial, and speckle is only a disturbance.}.

It may be also instructive to rewrite Eq.(\ref{eq:multiplicative}) as
\begin{equation}
    y = x+x(n-1) = x+n'
    \label{eq:additive}
\end{equation}
so as to obtain an {\em additive signal-dependent noise model}.
Indeed, the new noise term, $n'$, has zero mean but, unlike in more conventional systems, a variance that depends on the signal itself.
So, if we consider a homogeneous region, with constant signal level $x_0$, the signal will be affected by an additive noise term with the very same standard deviation, $x_0$,
such to have a local signal-to-noise ratio (SNR) of 0 dB, much smaller than the typical SNR's observed for optical images.
Needless to say, such a strong noise can be highly disruptive for the extraction of useful information, masking altogether precious fine details of the imaged scene.
Hence, removing or at least reducing speckle is of paramount importance for the success of SAR image processing applications.

A first and foremost way to reduce speckle intensity is multilooking, which consists in the incoherent averaging of multiple independent observations of the same stochastic signal.
The multilooked signal keeps obeying the multiplicative noise model of Eq.(\ref{eq:multiplicative}),
with the same signal component as before, but the noise term is now Gamma-distributed with pdf
\begin{equation}
    p(n) = \frac{L^L}{\Gamma(L)} n^{L-1} \exp(-L n) {\rm u}(n)
    \label{eq:gamma}
\end{equation}
with $L\geq 1$ representing the number of looks, and $\Gamma (\cdot)$ the gamma function.
In particular, due to the averaging, the noise mean keeps being unitary, $E[n] = 1$, but its variance reduces linearly with the number of looks, $\mbox{Var}[n]=1/L$,
thus providing much cleaner images.
Unfortunately, both frequency and spatial multilooking, easily implemented, cause a loss of spatial resolution, proportional itself to $L$,
which impairs significantly the value of the SAR image for the most advanced and demanding applications.
Temporal multilooking, on the contrary, preserves spatial resolution, but relies on hypotheses of signal ergodicity that are rarely met.
For these reasons, in the following we will mostly focus on single-look images, $L=1$,
which are the most valuable for applications, due to their high resolution, and represent also the most scientifically interesting condition.

\begin{figure}[t]
\includegraphics[width=0.225\textwidth]{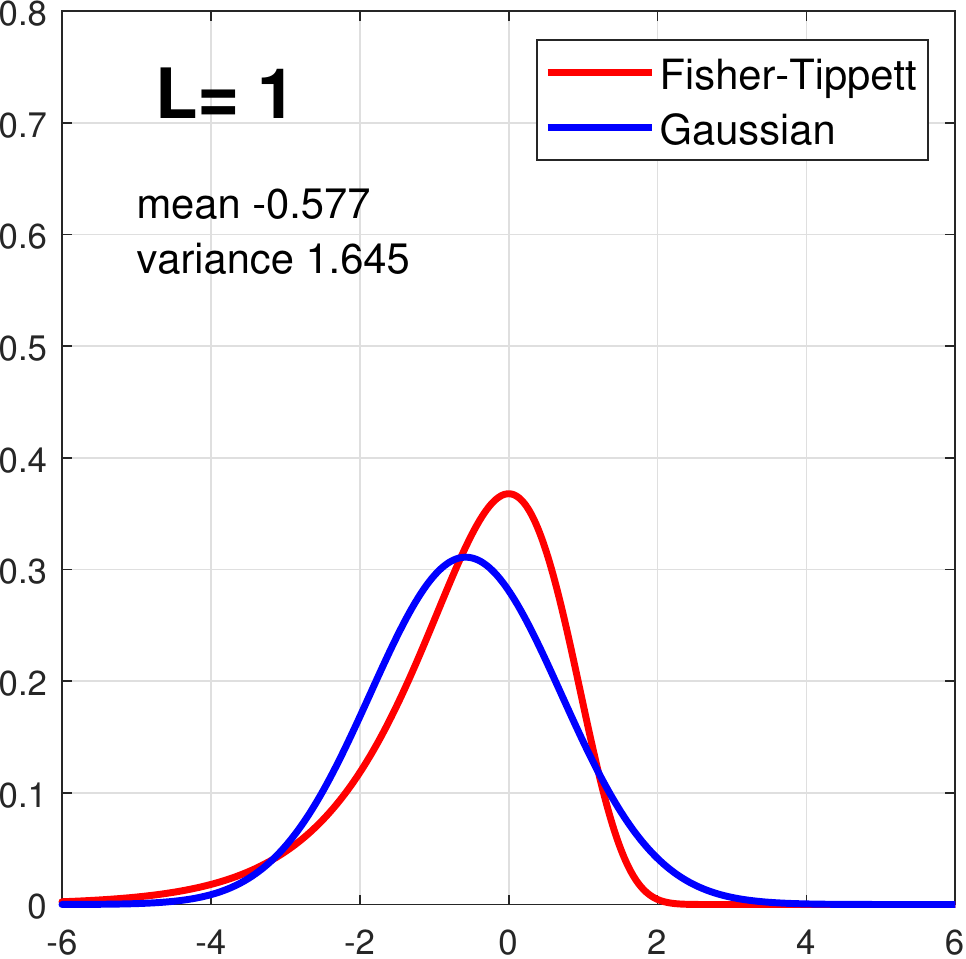} \hspace{2mm}
\includegraphics[width=0.225\textwidth]{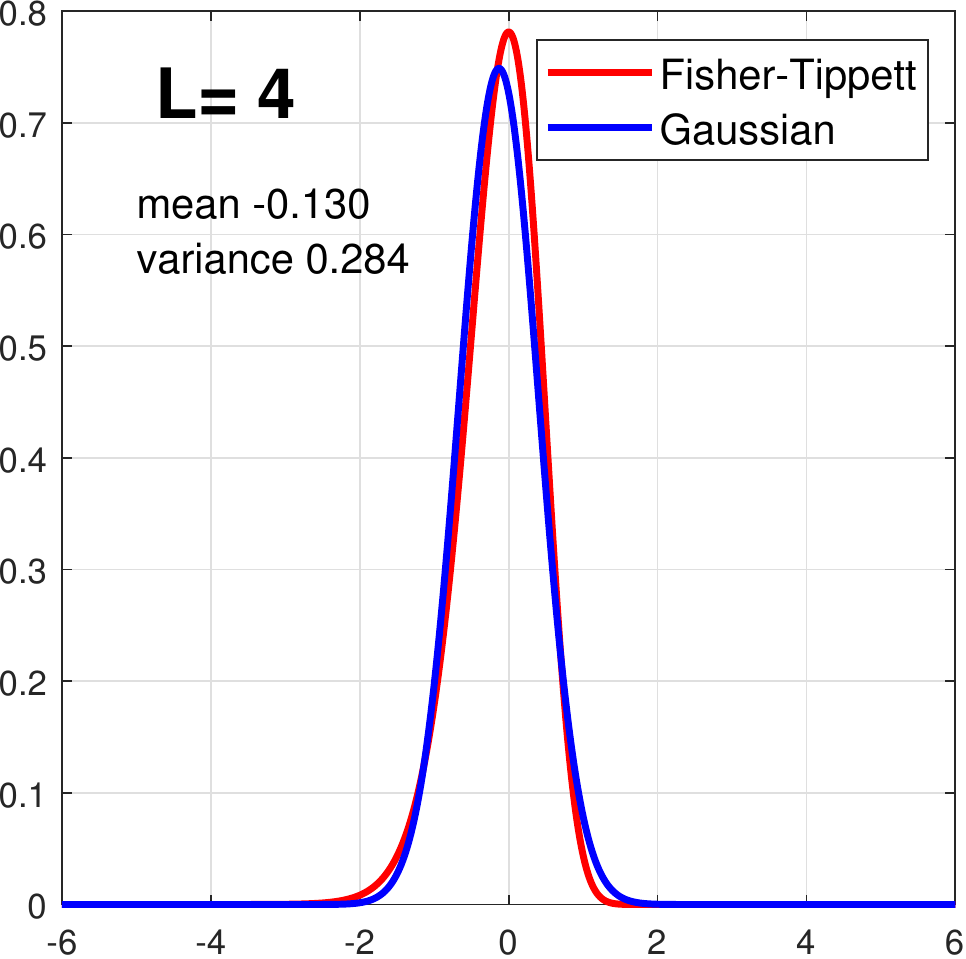}
\caption{Comparing Fisher-Tippet to Gaussian pdfs. Left: $L=1$. Right: $L=4$.}
\label{fig:pdfs}
\end{figure}

Dealing with multiplicative noise may be challenging and, on the other hand,
there is a vast literature on the removal of additive (signal-independent) noise, with a wealth of ideas, methods and algorithms that are definitely worth considering.
For these reasons, a common approach to despeckling, called the homomorphic approach,
is to take the log of the signal, so as to convert the problem into an additive one, perform additive noise removal, and then take the exp of the result to go back to the original domain.
Taking the log of Eq. \ref{eq:multiplicative} we obtain a plain {\em additive noise model}
\begin{equation}
    \ty = \tx + \tn
    \label{eq:homomorphic}
\end{equation}
where the log-domain noise, $\tn$, follows a Fisher-Tippett distribution \cite{pierce2002statistical}
\begin{equation}
    p(\tn) = \frac{L^L}{\Gamma (L)}  e^{L \tn} \exp (-L e^{\tn}) {\rm u(n)}
    \label{eq:fisher-tippett}
\end{equation}
with mean $E[\tn] = \psi(L)-\log(L)$, and variance $\mbox{Var}[\tilde{n}] = \psi (1,L)$,
where $\psi(\cdot)$ and $\psi(\cdot, L)$ are the digamma function and the polygamma function of order $L$, respectively.
As desired, neither mean nor variance depend on the signal.
However, the mean is non-zero, and will not be removed by any denoising algorithm, contributing a bias that must be compensated explicitly before going back to the original domain.
The desire to exploit methods developed for the popular additive white Gaussian noise (AWGN) case
motivates some researchers to approximate $\tn$ as non-zero mean Gaussian noise.
Fig.\ref{fig:pdfs} compares some Fisher-Tippet distributions with the Gaussian distributions having the same mean and variance.
While for the case $L=4$ a reasonably good match can be seen,
this is certainly not the case for the most interesting condition of $L=1$, for which the Gaussian approximation is certainly unsuitable.
Moreover, it should be noted that the SAR speckle is not spatially white.
In fact, to prevent strong scatterers to leak signal in neighboring cells,
SAR systems adopt a filtering technique, called apodization, which reduces such leakages, due to antenna sidelobes, at the cost of a loss of resolution, related to the antenna main lobe.
As a side effect, a moderate spatial correlation arises.
This is obviously neglected by AWGN-oriented methods, but may well be exploited to improve SAR despeckling performance \cite{lapini2014blind}.

\section{Primer on deep learning} \label{sec:primer_deep}

Recent years have witnessed the rise of machine learning methods to address a number of problems in the image processing and computer vision fields. Such data-driven methods typically rely on deep neural networks to act as universal function approximators, using some training data to learn a mapping between an input and the corresponding desired output. Within this general framework, a distinction must be made between supervised, self-supervised and unsupervised learning methods. Supervised methods are the most common and rely on having access to labeled data, i.e., data for which both the input and the desired ground truth output, e.g., a class label in a classification problem or a clean image in a denoising one, are available. While supervised training of deep neural networks can provide excellent results due to its ability to learn very complex mappings, it is ultimately limited by the need for large amounts of data with accompanying ground truth labels. Unsupervised methods, instead, do not rely on ground truth labels and seek to uncover latent properties of data by analyzing their features. Self-supervised learning can be regarded as a special case of both supervised and unsupervised learning in which the ground truth labels are not available, but it is still possible to learn a mapping to an (unknown) desired output, like in the supervised setting, by generating labeling information from the data themselves.

A deep neural network essentially amounts to a sequence of linear vector operations, parameterized by some weights, and interleaved by non-linear functions. Training a neural network amounts to finding the values of its parameters that minimize the loss function specified by the setting. While all neural networks can approximate arbitrary continuous and differentiable functions, different architectures provide different priors and inductive biases that can be useful for specific applications. In this sense, the convolutional neural network (CNN) proved to be valuable for problems concerned with visual data, such as images, video, and more. A CNN is composed by a stack of layers, each implementing a filter bank and a non-linear activation function. The input $\mathbf{Z}^{(l)} \in \mathbb{R}^{H^{(l)} \times W^{(l)} \times F^{(l)}}$ to the $l$-th convolutional layer is a stack of $F^{(l)}$ feature maps with spatial dimensions $H^{(l)} \times W^{(l)}$. An output feature map is computed by means of spatial convolution with a kernel $\mathbf{\Phi}_{i,f}^{(l)} \in \mathbb{R}^{K\times K}$ for each of the input feature maps, as indexed by $i$, and aggregation over all of them. This is then repeated to generate the desired number of output feature maps. In formulas:
\begin{align*}
    \mathbf{Z}^{(l+1)}_f = \sigma \left( \sum_{i=1}^{F^{(l)}} \left[ \mathbf{\Phi}_{i,f}^{(l)} * \mathbf{Z}^{(l)}_i \right] \right), \quad f=1,\dots,F^{(l+1)}.
\end{align*}
where $\sigma$ is a non-linear activation function and $*$ the convolution operator.

CNNs have been very successful at processing visual data because they enforce, by design, some priors that are true for natural images. In particular, the kernels have small spatial extent $K$ and this induces a localized receptive field, whereby the output of a layer at a given pixel is only affected by neighboring ones. The fact that the same kernel weights are reused over the whole image captures the stationarity property, where the characteristics of a feature do not depend on its spatial location. The small kernels and the spatial reuse due to the convolution operation also conveniently reduce the number of trainable parameters with respect to fully-connected neural networks, increasing efficiency and reducing the risk of overfitting. Finally, stacking many such layers creates a compositional representation, i.e., a hierarchy of features where higher level ones can be built by combining lower level ones. 

Another common architecture in visual problems is the Generative Adversarial Network (GAN) \cite{goodfellow2014generative}. GANs are usually employed to learn to generate new data samples whose distribution approximates the distribution of the training data. The architecture of a GAN is composed of two networks: a generator $G$ that learns to capture the distribution of the training data, and a discriminator $D$ that learns to distinguish between real training data and samples generated by $G$. The key insight of GANs is treating the training process as a game between these two networks, where the goal of $G$ is to generate data samples that can fool $D$ and the goal of $D$ is to tell apart the real and fake samples. GANs are commonly employed in image restoration problems in order to enforce that the data distribution of the restored images matches the one of the clean images.

Finally, the minimization of the loss function required to train deep neural networks is performed by means of first-order optimization methods such as stochastic gradient descent (or momentum-accelerated variants \cite{kingma2014adam}). Training is a computationally-intensive process that greatly benefits from the use of high-parallel hardware architectures such as Graphics Processing Units (GPUs).

\section{SAR Data} \label{sec:SAR_data}
This section discusses the key component of any data-driven algorithm, i.e., the data. First, we 
delve into the challenges that data-driven despeckling techniques face, discussing the various data exploitation options. Then, we analyze which datasets are available to tackle the SAR despeckling problem, emphasizing whether they can be retrieved from the public domain or  there are inherent difficulties in accessing them. In this treatment, we also discuss how the available data have been used by the remote sensing community so far, highlighting a lack of reproducibility or standardized testing procedures for fair evaluation.

\begin{figure*}
    \centering
    \includegraphics[width=0.36\textwidth]{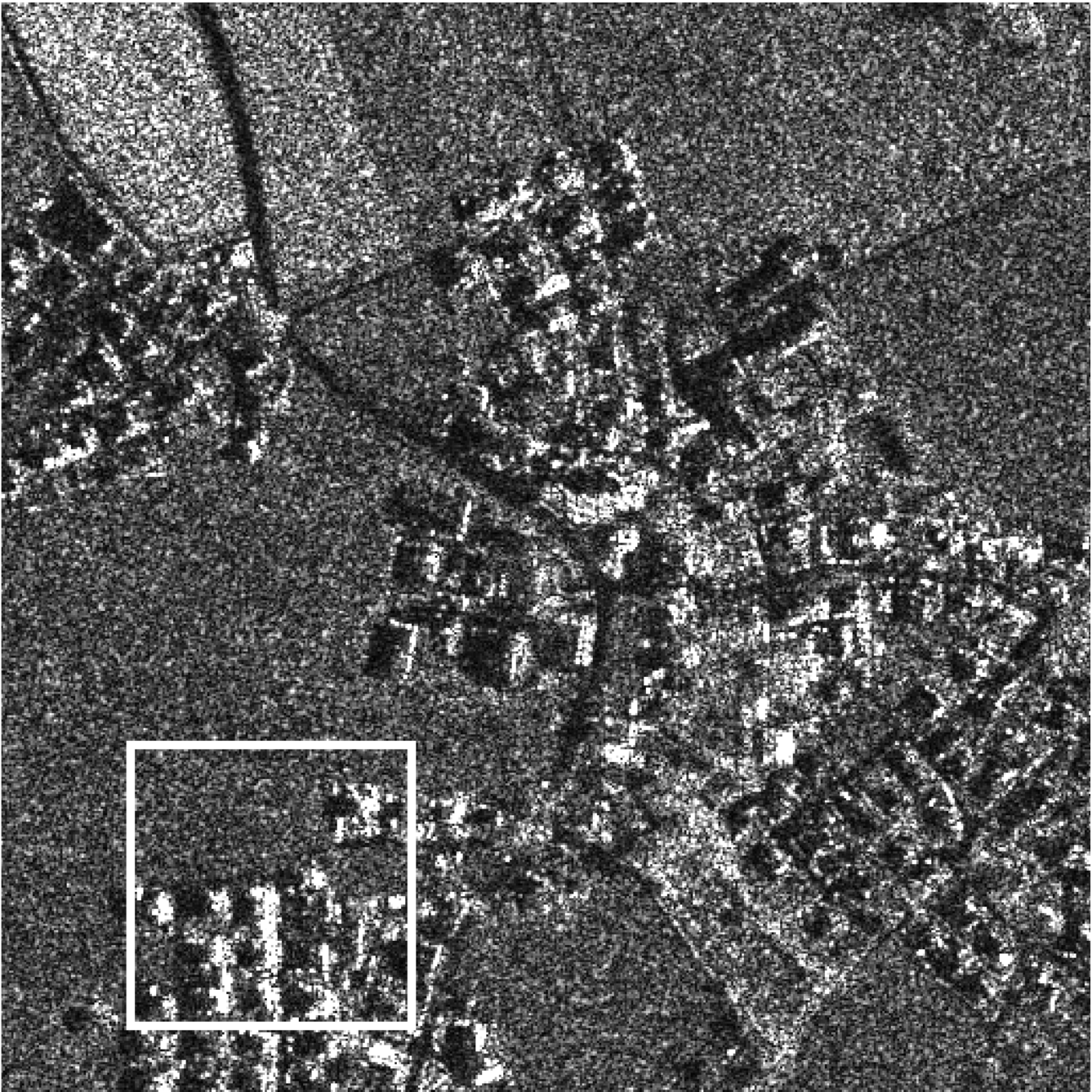}
    \includegraphics[width=0.36\textwidth]{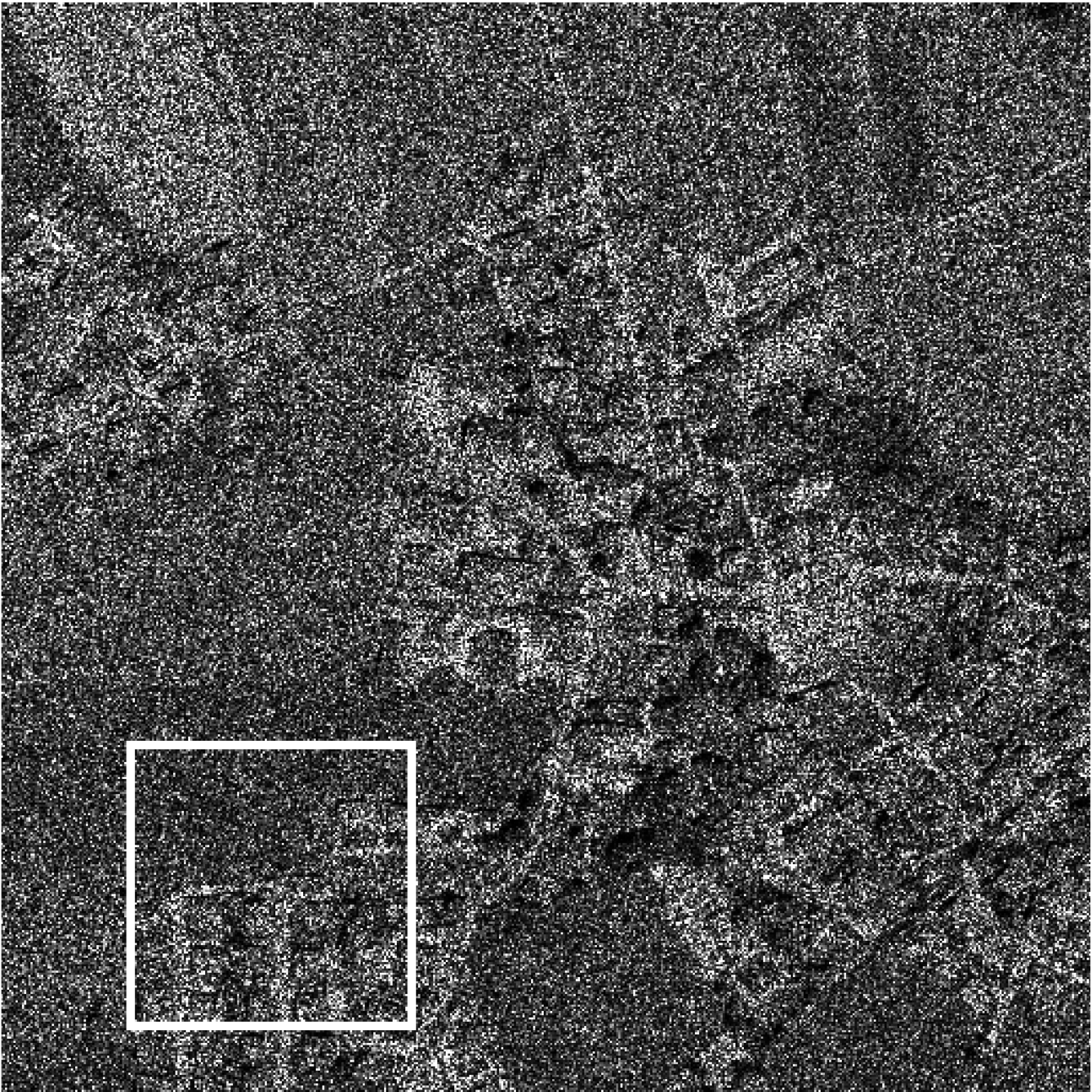} \\ \vspace{1mm}
    \includegraphics[width=0.36\textwidth]{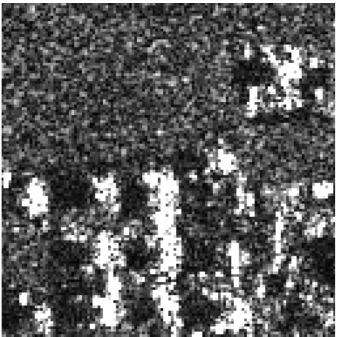}
    \includegraphics[width=0.36\textwidth]{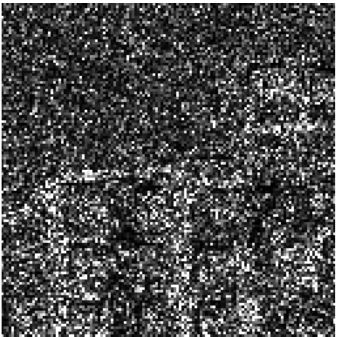}
    \caption{Drawbacks of the fully simulated approach to training.
    Top-right, the TerraSAR-X image of Fig.\ref{fig:speckle}, with a clip highlighted by a white box.
    Top-left, the synthetic SAR image simulated by injecting single-look speckle on the co-registered gray-scale optical image. 
    The gray levels of corresponding regions are unrelated, due to the different imaging mechanisms, but this is immaterial.
    Real problems are highlighted by the enlarged clips shown in the bottom row.
    The urban areas are completely different in terms of spatial structure: double-reflection lines and shadows of the SAR image are lost in the simulated image and corrupted by noise.
    Even in the homogeneous area, the speckle is clearly correlated in the real SAR image, not so in the simulation.}
    \label{fig:fully_simulated}
\end{figure*}

\begin{figure*}
    \centering
    \includegraphics[width=0.36\textwidth]{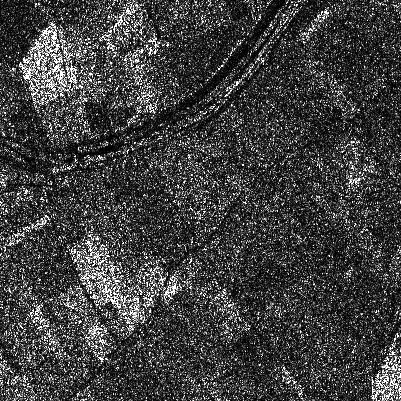}
    \includegraphics[width=0.36\textwidth]{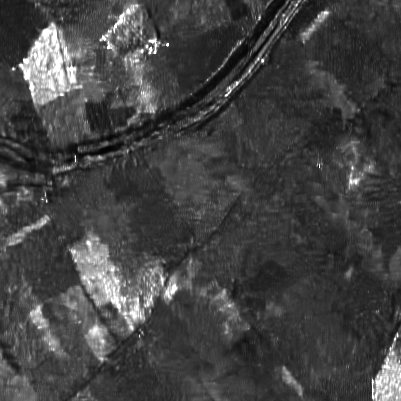}\\ \vspace{1mm}
    \includegraphics[width=0.36\textwidth]{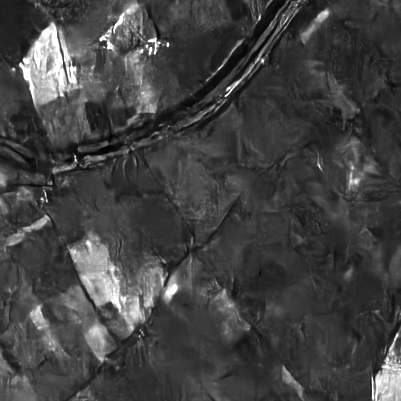}
    \includegraphics[width=0.36\textwidth]{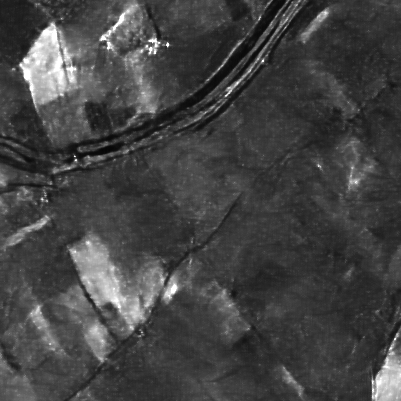}
    \caption{Example of domain gap artifacting due to supervised training with synthetic speckle. Top-left: single-look noisy image. Top-right: model-based denoising (SAR-BM3D; \cite{Parrilli2012SARBM3D}). Bottom-left: supervised denoising (ID-CNN \cite{Wang2017SAR}). Bottom-right: self-supervised denoising (Speckle2Void \cite{Bordone2020speckle2void}). Notice the cartoon-like edges and hallucinations in the flat regions produced by the supervised method. Image from DLR TerraSAR-X.}
    \label{fig:artifacts}
\end{figure*}

\subsection{Data usage}

The success of the supervised learning setting in problems like image classification, segmentation, etc., has quickly led to the development of supervised denoisers \cite{zhang2017beyond,mao2016image,tai2017memnet,liu2018non,plotz2018neural,valsesia2020deep}, i.e. CNNs learning a mapping from noisy to denoised images using clean images of the same class as ground truth in the training process. This setup is appealing in the classical setting of  additive Gaussian noise removal because it is not difficult to retrieve a large quantity of virtually noiseless images (e.g., high-quality photos with very low levels of camera noise) and create an arbitrary number of noisy images by sampling realizations of Gaussian random variables. It is only natural that several works in the SAR despeckling literature tried to replicate this setting. However, any SAR despeckling technique must face the fact that no ``clean SAR images'' exist.

As discussed in Sec. \ref{sec:primer_speckle}, it is not possible to generate clean references by multilooking, as this process degrades the spatial resolution. Therefore, the literature addressing the despeckling problem through supervised deep learning (in Secs. \ref{sec:supervised_architectures} and \ref{sec:supervised_training}) has essentially presented two methods to generate reference images to be used as ground truth: synthetic speckle generation and multitemporal fusion. Each of these has its own advantages and disadvantages, which we are now going to discuss.

The \textit{synthetic speckle generation} approach starts from optical images, either satellite data or even plain photographs (e.g., the BSD500 photo dataset \cite{MartinFTM01} is commonly used), where the amount of noise can be considered negligible. It then uses a model of speckle, such as the one in Eq.\eqref{eq:multiplicative}, and a statistical characterization of the speckle, e.g. as in \eqref{eq:gamma}, to sample realizations of the speckle process and superimpose it on the clean optical data. This procedure is a simple means to generate data that, at a first glance, resemble speckled SAR images and can be used to train denoisers for multiplicative instead of additive noise. However, this simplicity comes at the cost of several disadvantages. 
First and foremost, while synthetically speckled optical images may vaguely resemble actual SAR images, due the different properties of materials they match neither their spatial structure nor their radiometric statistics, which is not surprising given the completely different nature of the two imaging mechanisms.
In particular, the existence of strong reflectors is not modelled by using optical images; moreover, priors on texture patterns and edges learned from optical data may not match the characteristics of actual SAR data. Fig. \ref{fig:fully_simulated} shows an example of the differences between a real SAR image and one with synthetically generated speckle. This mismatch is the source of a problem often referred to as \textit{domain gap} \cite{gretton2009covariate}, whereby the features of the test set (real SAR images) may differ from those of the training set, and induce anomalous inference behavior. This can appear as artifacts, hallucinations of patterns that were prevalent in the training set, or oversmoothing. Fig.\ref{fig:artifacts} shows an example where a CNN trained on synthetically speckled optical images tends to produce a cartoon-like despeckled image, either oversmoothing regions or hallucinating patterns in supposedly flat regions. The domain gap is a serious problem that should not be disregarded, as it reduces the confidence of the final users on the generated products.
Second, the denoiser will only be as good as the model it relies on; there are several details that are overlooked by simple models such as those presented in Sec.\ref{sec:primer_speckle}. An important example is the assumption that the speckle process is spatially uncorrelated. This is an almost universally used assumption, but it is hardly satisfied in practice, as it has been observed that the point spread function of the SAR focusing algorithms generates spatially correlated speckle \cite{lapini2014blind}.

\begin{table*}
\centering{
\addtolength{\tabcolsep}{-4pt}
\caption{SAR Data Availability}
\label{table:datasets}
\begin{tabular}{lccll}
\textbf{Platform} & \textbf{Year} & \textbf{Description} & \textbf{Availability} & \textbf{Link} \\ \hline \hline
 Sentinel-1 & 2014 - present  & \cite{sentinel-1} & available upon registration & \url{https://scihub.copernicus.eu/} \\ \hline
 TerraSar-X & 2007 - present & \cite{terrasar-x} & available upon registration & \url{https://tpm-ds.eo.esa.int/oads/access/collection/TerraSAR-X} \\ \hline
& & & limited availability &  \\ 
 Cosmo-SkyMed & 2008 - present &\cite{cosmoskymed}&only for users in ESA Member States, &\url{https://tpm-ds.eo.esa.int/oads/access/collection/CosmoSkyMed}\\
&&&European Commission Member States,&\\
&&&Canada, Africa and China&\\\hline
&  &  & limited availability & \\ 
 RadarSAT-2 & 2008 - 2019 &  \cite{radarsat2} &only for users in ESA Member States & \url{https://tpm-ds.eo.esa.int/oads/access/collection/Radarsat-2}\\ 
&&&and European Commission Member States&\\\hline
 AIRSAR & 1990 - 2004 & \cite{airsar} & available upon registration & \url{https://asf.alaska.edu/data-sets/sar-data-sets/airsar/} \\ \hline
 ALOS-PALSAR & 2006 - 2011& \cite{alos} & available upon registration  & \url{https://alos-palsar-ds.eo.esa.int/oads/access/}$^*$\\ 
 &  &&  & \url{https://asf.alaska.edu/data-sets/sar-data-sets/alos-palsar/}$^\dagger$\\\hline
 ERS-2&1995 - 2011&\cite{ers-2}&available upon registration&\url{https://asf.alaska.edu/data-sets/sar-data-sets/ers-2/}\\
\hline
\end{tabular}%
}
\begin{flushleft}
$^*$Spatial coverage: ADEN zone (Europe, Africa and the Middle East) plus some worldwide products received from JAXA.\\
$^\dagger$Spatial coverage: the Americas, Antartica and selected global sites.
\end{flushleft}
\end{table*}

The \textit{multitemporal fusion} approach uses a stack of actual SAR images acquired at different temporal istants and exploits the temporal incoherence of the speckle to effectively suppress it and generate a clean ground truth. The clear advantage is that a denoiser trained with this approach would not suffer from the aforementioned domain gap problem, thus providing high-quality and reliable results. However, the challenges associated with this technique lie in the dataset creation phase. First of all, one needs to access to a large enough repository of multitemporal images. This can be challenging in some cases due to data distribution policies, as described in Sec.\ref{sec:data_avail}. Even if data are available, the multiple versions of the same scene must be accurately registered to one another, correcting for any small disparity or pitch variations of the sensing platform. This registration process may itself affect the quality of the data by changing their statistical properties. Finally, a major limitation of multitemporal fusion is represented by change in the scene content as a result of human activity, natural phenomena, etc. resulting in poor accuracy in the areas affected by it. 

It is worth mentioning that the most recent approaches to tackle the ground truth problem rely on avoiding the need for ground truth altogether. The self-supervised learning approaches we discuss in Sec.\ref{sec:selfsupervised} learn to perform the despeckling task by only using noisy images. This is an interesting development as learning from real data allows to avoid any domain gap and fully exploit the data. However, more work is needed on these emerging approaches as they still require either multitemporal data or careful modelling of the prior knowledge about the images and speckle process.

\subsection{Data availability} \label{sec:data_avail}

Deep learning methods require very large amounts of data to effectively train a network. For this reason, the availability of data is an issue of primary importance in this context. In the past, the availability of SAR data was very limited: some datasets were only available for certain areas or only upon payment, while in other cases their access was limited due to confidentiality issues. Only recently the situation changed and some large SAR datasets are now widely available for free. This novel open data policy was first introduced by the European Spatial Agency with the launch of the Copernicus programme in 2014. As a consequence, the imagery acquired by several SAR sensors can now be easily found online. In Table \ref{table:datasets}, we list the main repositories of SAR images that are publicly available and can be accessed online, also describing their main characteristics as well as their data release policies. 

As discussed in the previous section, some deep learning methods for SAR despeckling are trained on multitemporal stacks of SAR images, i.e., SAR images of the same scene acquired at different times. The creation of such type of datasets is critical. In particular, one of the main difficulties is to retrieve a large enough number of multitemporal images. Some of the data sources presented in Table \ref{table:datasets}, such as Sentinel-1 and TerraSAR-X, provide acquisitions of the same scenes at different moments in time and allow to create large multitemporal datasets, which can be effectively used for training. For example, Sentinel-1 has a revisit time of 6 days, which allows for fairly dense temporal sampling.

Moreover, several deep learning methods for SAR despeckling use optical images to train the model in a supervised fashion, as explained in the previous section. In this case, general purpose datasets used for visual problems, such as the BSD500 [41], can be employed. Aerial images are also widely used for training SAR despeckling methods, such as those of the Google Maps dataset \cite{isola2017image}, the UC Merced Land Use dataset \cite{yang2010bag} and the NWPU dataset \cite{Cheng_2017}. The images contained in such datasets are more similar to actual SAR images, therefore including them in the training set will help to learn priors that can better match the characteristics of SAR data and will result in a reduced domain gap. 

In this case, general purpose datasets used for visual problems, such as the BSD500 [41], can be employed. Aerial images are also widely used for training SAR despeckling methods, such as those of the

Even though today many SAR datasets are available and can be used for training and testing, a standardized procedure for the experimental evaluation of the despeckling methods is still lacking: every work considers a different dataset and, even when the dataset is the same, the partition between train and test set can change. Moreover, it is often the case that the datasets used for training and testing are not clearly described. This severely hinders the reproducibility of the results and precludes the possibility of a fair comparison between different methods. These issues represent a serious problem for the advancement of the research in this field, since the lack of a rigorous assessment of the performance creates confusion and makes it impossible to discern the valid contributions that provide advances with respect to the state of the art.  
For these reasons, there is a strong need for standardized datasets with predefined test and train splits that can lead to a fair comparative evaluation and reproducible results. This is the only way the research in this field can continue to make progress, and this aspect is discussed further in Sec. \ref{sec:future_directions}.

\section{Supervised models: architectures}
\label{sec:supervised_architectures}

We now review despeckling methods based on {\em supervised} deep learning models, focusing first on architectures (this Section) and then on training (Section \ref{sec:supervised_training}).
Methods are divided in two large families depending on their architecture.
The first one comprises ``direct'' methods, based on a plain deep network that accepts as input the noisy image and outputs its denoised version.
In the second family we cluster more elaborate methods, in which deep networks are used in the context of a larger despeckling procedure, for example, to estimate some parameters of a model-based denoising engine.
For direct methods, special attention will be devoted to the fundamental choice of the noise model, explaining its tight relationship with the filtering architecture.
A summary of the analyzed methods with their main features is reported in Tab.\ref{table:supervised}.

\subsection{Direct DL-based despeckling}

\begin{figure}
    \centering
    \includegraphics[width=0.45\textwidth]{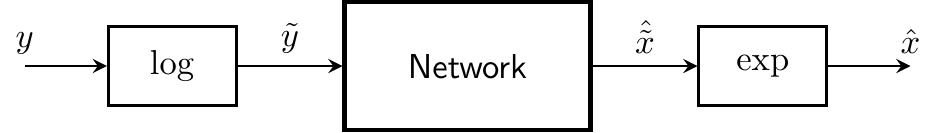}\\ \vspace{5mm}
    \includegraphics[width=0.45\textwidth]{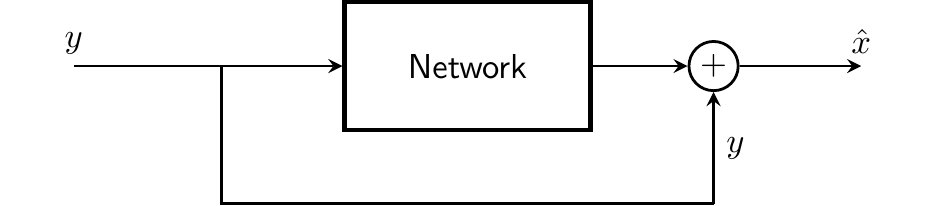}\\ \vspace{5mm}
    \includegraphics[width=0.45\textwidth]{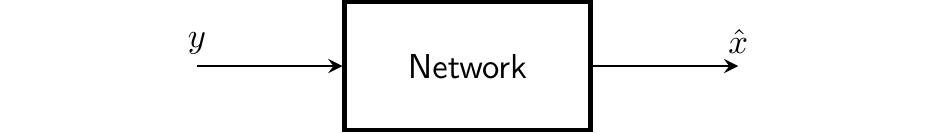}
    \caption{Network architecture designs. Top to bottom: direct DL-based despeckling in the log domain, direct DL-based despeckling with a residual architecture, direct DL-based despeckling with a plain architecture.}
    \label{fig:supervised}
\end{figure}

Most methods proposed in the literature belong to this family and, even in the context of direct filtering, display a large variety of architectural solutions.
So, we further divide them depending on two fundamental choices (overviewed in Fig. \ref{fig:supervised}): working in the original or in the log-domain, and using a residual or a plain network.
Indeed, these choices are related, more or less explicitly, to basic assumptions on the noise modeling.
\begin{enumerate}
\item   {\it Log-domain methods:}
        by taking the log of the input image, the noise is converted from multiplicative to additive, irrespective of whether a residual architecture is used or not.
        Moreover, the variance of the noise is stabilized, which fully justifies the adoption of fixed denoisers.
        This allows one to tap into the huge reservoir of ideas and methods developed for the restoration of AWGN images.
        However, one should be aware that the log-domain noise is neither white nor Gaussian, but weakly correlated Fisher-Tippett (see Section \ref{sec:primer_speckle}).
\item   {\it Original-domain residual architectures:}
        If a residual architecture is used in the original domain, the input image is regarded as the sum of a clean image (ideally, the output of the networks) and noise.
        Therefore an additive noise model is still used, but now the noise has signal-dependent variance: it is more intense in regions with high reflectivity, and less intense in regions with low reflectivity.
        This may not be a problem if the network operates on small homogeneous patches of the image.
        On the contrary, heterogeneous patches will be characterized by additive noise of spatially-varying intensity, with unpredictable effects.
\item   {\it Original-domain no-residual architectures:}
        In this case, the network operates on the image as is, hence it has to deal with a truly multiplicative noise.
        Note that also the method proposed in \cite{Wang2017SAR} fits here, since it uses a residual architecture based on a ratio of images rather than a difference, fully consistent with the multiplicative model.
\end{enumerate}
We now analyze in turn these groups of methods.
Unless explicitly stated, we consider the fully developed speckle model to hold, and refer to the most challenging case of single-look speckle.

\newcommand{\ru}{\rule{0mm}{3.2mm}}
\setlength{\tabcolsep}{2mm}
\begin{table*}
\caption{Some relevant deep learning-based despeckling methods with their main features}
\label{table:supervised}
\begin{tabular}{cllllll}
		\ru   \textbf{Ref.}                           & \textbf{Acronym}       & \textbf{Keywords}                  & \textbf{Use of DL }       & \textbf{Noise model}    & \textbf{Training set}         &  \textbf{Code}   \\ \hline \hline
		\ru   \cite{Chierchia2017SARCNN}     & SAR-CNN       & residual@log              & direct  & additive       & 25-look COSMO        & \url{grip-unina.github.io/SAR-CNN/} \\ \hline
		\ru   \cite{yang2019sar}             & MuLoG-CNN     & MuLoG w/DnCNN             & direct  & additive       & N-look Sentinel      &         \\ \hline
		\ru   \cite{dalsasso2020sar}         & MuLoG-CNN     & pretrained vs. training   & direct  & additive       & N-look Sentinel      & \url{gitlab.telecom-paris.fr/ring/SAR-CNN} \\ \hline
		\ru   \cite{pan2019filter}           & MuLoG+FFDNet  & MuLoG w/FFDNet            & direct  & additive       &                      &         \\ \hline
		\ru   \cite{yue2018sar}              & DNN           & signal KLD loss           & direct  & additive       &                      &         \\ \hline
		\ru   \cite{Zhang2018learning}       & SAR-DRN       & dilated convolutions      & direct  & add-sig-dep    & UCMerced             & \url{github.com/qzhang95/SAR-DRN} \\ \hline
		\ru   \cite{Li2018HDRANet}           & HDRANet       & attention modules         & direct  & add-sig-dep    & UCMerced             &         \\ \hline
		\ru   \cite{Gui2018SAR}              & SAR-DDCN      & dense+dilated             & direct  & add-sig-dep    & UCMerced             &         \\ \hline
		\ru   \cite{ZhangJing2019SIDU}       & MCN-WF        & blockwise dense           & direct  & add-sig-dep    & UCMerced             &         \\ \hline
		\ru   \cite{Lattari2019Deep}         & n/a           & U-Net                     & direct  & add-sig-dep    & N-look Sentinel      &         \\ \hline
		\ru   \cite{gu2017residual}          & REDNET        & REDNet                    & direct  & add-sig-dep    & BSD                  &         \\ \hline
		\ru   \cite{Wang2017SAR}             & ID-CNN        & residual w/ratio          & direct  & multiplicative & UCID, BSD, Gmaps     & \url{github.com/XwK-P/ID-CNN} \\ \hline
		\ru   \cite{zeng2018speckle}         & n/a           & ELU nonlinearity          & direct  & multiplicative &                     &         \\ \hline
		\ru   \cite{bai2018fractional}       & FID-CNN       & fractional TV             & direct  & multiplicative &                      &         \\ \hline
		\ru   \cite{Wang2017generative}      & ID-GAN        & GAN+perceptual loss       & direct  & multiplicative & UCID, BSD, Gmaps     &         \\ \hline
		\ru   \cite{liu2020gan}              & n/a           & GAN+TV loss               & direct  & multiplicative & BSD,UCMerced         &         \\ \hline
		\ru   \cite{liu2020spatial}          & STD-CNN       & attention modules         & direct  & multiplicative & BSD                  &         \\ \hline
		\ru   \cite{ferraioli2019novel}      & n/a           & speckle KLD loss          & direct  & multiplicative & UCID, BSD, Gmaps     &         \\ \hline
		\ru   \cite{vitale2019new}           & KL-DNN        & speckle KLD loss          & direct  & multiplicative & UCID, BSD, Gmaps     &         \\ \hline
		\ru   \cite{vitale2020multi}         & MONet         & speckle KLD loss          & direct  & multiplicative & UCMerced             &         \\ \hline
		\ru   \cite{tang2019sar}             & MLP           & MLP                       & direct  & multiplicative & 52-look TerraSAR-X   &         \\ \hline
		\ru   \cite{zheng2018sar}            & n/a           & residual w/ratio          & direct  & multiplicative &                      &         \\ \hline
		\ru   \cite{cozzolino2019nonlocal}   & CNN-NLM       & nonlocal+DL               & blended          & multiplicative & 25-look COSMO        & \url{github.com/davin11/CNN-NLM} \\ \hline
		\ru   \cite{Cozzolino2020Nonlocal}   & CNN-NLM       & nonlocal+DL, N$^3$ layers & blended          & multiplicative & 25-look COSMO        & \url{github.com/davin11/CNN-NLM} \\ \hline
		\ru   \cite{denis2019patches}        & n/a           & MuLoG+nonlocal            & blended          & additive       & n/a                  &         \\ \hline
		\ru   \cite{zeng2019computational}   & SAR-NN3D      & nonlocal+DL               & blended          & unspecified    & pre-trained          &         \\ \hline
		\ru   \cite{shen2020sar}             & SAR-RDCP      & MAP w/ CNN prior          & blended          & add-sig-dep    & UCMerced             &         \\ \hline
		\ru   \cite{liu2019convolutional}    & n/a           & guided fusion             & blended          & additive       & BSD                  &         \\ \hline
		\ru   \cite{gu2020two}               & n/a           & texture map               & blended          & multiplicative & NWPU                 &         \\ \hline
		\ru   \cite{liu2020sar}              & n/a           & shearlet                  & blended          & additive       &                     &         \\ \hline
		\ru   \cite{zhou2019deep}            & MSR-Net       & multiresolution           & blended          & multiplicative & UCMerced             &         \\ \hline
\end{tabular}

\vspace{2mm}
{The acronyms are those used by the authors in the original papers.}
\end{table*}

\subsubsection{Log-domain methods}

SAR-CNN \cite{Chierchia2017SARCNN}, one of the first CNN-based SAR despeckling methods proposed in the literature, often considered as baseline, follows this approach.
The log of the input image feeds a 17-layer CNN, which is a straightforward adaptation of the DnCNN denoiser proposed in \cite{zhang2017beyond} for the AWGN case, based on a residual architecture \cite{He2016residual}.
Therefore, the CNN extracts the log-domain noise, which is then subtracted from the original image before compensating for the non-zero mean and taking the exp of the result.
In SAR-CNN the network is trained on simulated single-look SAR images.
However, to ensure a better fidelity to the actual statistics of SAR signal and speckle, it is retrained on real SAR data, using multilooked images as approximate clean references.

The reliance on AWGN filters is made fully explicit in methods based on the MuLoG \cite{deledalle2017mulog} paradigm.
Considering the scarcity of reliable training data for SAR image despeckling,
the goal of MuLoG is to use AWGN denoisers just as they are, with a standard adaptation procedure to fit them to the Fisher-Tippett distribution of log-transformed SAR speckle.
Therefore, the approach works both with conventional and DL-based denoisers, the latter trained on unlimited AWGN data.
In \cite{yang2019sar} MuLoG-based despeckling is performed using both a model-based AWGN denoiser (BM3D \cite{dabov2007image}) and the pre-trained DnCNN.
Experiments prove the importance of the adaptation phase (that is, the unsuitability of straight AWGN denoisers for log-domain SAR data) and the superiority of the deep learning denoiser.
Yet, further analyses carried out in \cite{dalsasso2020sar} show that the performance remains below the level of SAR-CNN (a slightly improved version implemented by the authors),
underlining the importance of training or at least fine-tuning the network on real SAR data.
The MuLoG approach is also followed in \cite{pan2019filter}, where the DnCNN engine is replaced with the newer and more effective FFDNet \cite{zhang2017ffdnet}, with some beneficial effects on performance.

On the opposite side of the spectrum, \cite{yue2018sar} models not only the speckle but also the signal itself as a random process, so as to better take into account the homogeneous/heterogeneous nature of the observed cell.
Working in the log-domain, the pdf of the observed signal can be regarded as the result of a convolution between the pdfs of clean signal (unknown) and speckle.
By means of an elaborate procedure, a rather shallow CNN is trained to predict pdf and mean of the clean signal.
Experiments on synthetic data provide some support to this approach, but results on real-world SAR images are still inconclusive.

\subsubsection{Original-domain residual architectures}

This is a fairly large group of methods as the residual architecture is quite widespread.
The first method to adopt this setting, SAR-DRN \cite{Zhang2018learning}, is by now one of the most popular in the field, often used as a baseline.
An appealing feature is its lightweight architecture, with only seven convolutional layers.
This choice reduces training complexity without impairing performance, thanks to the use of dilated convolutions, which ensure that a large aggregate receptive field is obtained anyway.
In addition, skip connections are also used to implement residual blocks.
The network is trained exclusively on simulated data.
Although questionable, this is such a common trait that, from now on, we will only point out exceptions to this rule.
HDRANet, proposed in \cite{Li2018HDRANet}, also uses dilated convolutions and skip connections in a 7-layer architecture.
The main innovation consists in the introduction of attention modules, operating both in space and across channels.
Channel attention modules redefine the convolutional features through suitable (attention) weights that emphasize effective features and suppress useless ones.
Likewise, spatial attention modules focus on image regions which are more informative.
Even though ablation studies seem to support the importance of attention modules, in general, it is not obvious how spatial attention, in particular, helps achieving a better image restoration.

SAR-DDCN, proposed in \cite{Gui2018SAR} is conceptually similar to SAR-DRN, the main innovation being the introduction of two 5-layer dense blocks.
Dense connections \cite{huang2017densenet} are well known to allow a better propagation of features in very deep networks and hence reduce the vanishing gradient problem, a major issue for the training of very deep CNNs.
However, the network used here is only 12-layer deep.
Dense connections are used also in MCN-WF (multiconnection network with wavelet features) proposed in \cite{ZhangJing2019SIDU},
now in the context of a much deeper network comprising 32 layers.
Indeed, the explicit goal of the method, based on the analysis of previous literature, is to use a deeper architecture to extract more expressive features.
The computational load is reduced by means of simplified dense connections: 5-layer blocks are used as in \cite{Gui2018SAR}, but only the last layer in the block receives inputs from all preceding layers.
Then, the same structure is replicated at the block level, thus constructing a hierarchical multiconnection network.
A further expedient is to compute preliminarily a single-level wavelet transform of the image, so as to deal with a more compact 4-subband input,
and also to gain the freedom to use different losses for low and high frequencies.

A relatively deep network is proposed also in \cite{Lattari2019Deep}, inspired by the U-Net architecture \cite{ronneberger2015unet} proposed originally for image segmentation.
This is an encoder-decoder network: in the encoder part, the image is repeatedly subsampled to extract rich contextual features.
The decoder then expands the features back to the image size.
Moreover, to preserve image details, several skip connections link the two branches of the ``U'' at the same sampling level, so as to inject high-resolution details in the output.
The architectural choices are well supported by ablation studies.
It is worth underlining that the network is trained also on realistic SAR data, obtained by injecting simulated speckle on a deeply multilooked SAR image.
Although the speckle is simulated, the statistics of the clean SAR signal are preserved.
Skip connections are also the key idea of \cite{gu2017residual} where a 28-layer residual encoder-decoder network (REDNET) \cite{mao2016image} is used.
Again, the goal is to preserve image details and reduce the vanishing gradient problem, but features are now added, like in a residual network, rather than concatenated like in U-Net.

\subsubsection{Original-domain no-residual architectures}

ID-CNN \cite{Wang2017SAR} is one of the earliest CNN-based image despeckling methods.
The proposed network has a residual architecture but, unlike previous methods, it aims at estimating the noise content from the original-domain image.
Therefore, the denoised image is obtained by taking the ratio, rather than the difference, between the input image and the estimated speckle.
This approach makes full sense, considering the multiplicative nature of noise.
Of course, a pointwise ratio of images may easily produce outliers, in the presence of estimated noise values close to zero.
However, a \textit{tanh} nonlinearity layer placed right before the output performs a soft thresholding thus avoiding serious shortcomings.
Despite the good performance, no other methods followed this path, except for some trivial variations \cite{zeng2018speckle, bai2018fractional} of ID-CNN itself.
The network itself is quite standard, with eight convolutional layers, batch normalization and ReLU.

In \cite{Wang2017generative} the same authors propose ID-GAN, a despeckling method based on a generative adversarial net.
Although GANs are not plain CNNs, we consider this to be a direct DL-based method, because the actual despeckling engine is just the generator subnetwork.
The generator takes the noisy input as a seed and generates a new image which must appear virtually speckle-free to pass the scrutiny of the discriminator.
Eventually, the generator is trained with a composite loss, which includes not only the adversarial term, but also the usual $\ell_2$ loss, to ensure fidelity to the original image,
and the perceptual loss proposed in \cite{johnson2015perceptual}, based on deep features extracted by an independent pre-trained VGG16 \cite{simonyan2015verydeep} model.
Apart from this, the generator is quite standard, a symmetric 8-layer CNN with auto-encoder structure.
As in many other cases, the model (trained on synthetic data) is not available online, which is unfortunate given the difficulty of GAN training.
A similar method, with minor architectural variations, is proposed in \cite{liu2020gan} with a total variation (TV) loss in place of the perceptual loss.

A multiplicative noise model is adopted also in \cite{ferraioli2019novel, vitale2019new, vitale2020multi}, all papers by the same group of authors.
In the first two proposals, a 10-layer plain CNN is used, while in the multi-objective network (MONet) of \cite{vitale2020multi}, two-layer residual blocks are also considered, bringing the network to a total of 17 layers.
Rather than on architecture, the main focus here is on the loss function, which is crafted so as to capture the statistical peculiarities of SAR images and speckle noise.
We mention briefly also \cite{tang2019sar} and \cite{zheng2018sar} where 3-layer nets are used, a plain CNN and a multilevel perceptron, respectively, with results of limited interest.

\subsection{Model-based despeckling methods exploiting DL tools}

Methods of this family try to blend the model-based and data-driven approaches,
with the aim to exploit both the large body of knowledge and procedures accumulated on SAR despeckling over several decades  and the great potential of deep learning tools.

A perfect example of this blending is represented by CNN-NLM \cite{cozzolino2019nonlocal, Cozzolino2020Nonlocal}, where despeckling is carried out by nonlocal means, a simple and well-understood linear filtering algorithm.
The clean target pixel is estimated as a weighted average of neighboring noisy pixels, with weights that depend on the similarity between target and estimator.
In CNN-NLM the similarity metric is replaced by a suitably trained CNN.
The network takes as input a patch extracted from the original-domain image, and outputs a set of filter weights, adapted to the local image content.
In \cite{cozzolino2019nonlocal} a rather standard CNN is used with 12 convolutional layers,
while in \cite{Cozzolino2020Nonlocal} a 20-layer CNN is proposed which includes also two $N^3$ layers proposed in \cite{plotz2018neural} to exploit image self-similarities.
These layers associate the set of its $K$ nearest neighbors with each input feature, which can be exploited for subsequent nonlocal processing steps.
Training is both on synthetic data and on real multilooked SAR images, like in \cite{Chierchia2017SARCNN}.
Results are much better than those of conventional nonlocal methods, like PPB \cite{deledalle2009iterative},
which provides some hints on how the filter weights should be chosen given the underlying signal and the noise strength.
Moreover, the performance matches that of state-of-the-art CNN-based methods, which is quite interesting, considering that the filtering engine is fully linear.
The fact that, despite the non-additive nature of the noise, a linear filtering method can be competitive with highly nonlinear deep networks may deserve further studies.

Note that the interplay between nonlocal filtering and deep learning is the object of intense research for AWGN denoising, {\it e.g.,} \cite{lefkimmiatis2017nonlocal,wang2018nonlocal,plotz2018neural,cruz2018nonlocality,valsesia2020deep}.
A first exploratory work for SAR despeckling, inspired to \cite{cruz2018nonlocality}, is carried out in \cite{denis2019patches},
where the output image provided by the MuLoG approach with the DnCNN denoiser is eventually subject to a nonlocal refinement.
Similar patches, selected based on their similarity in the original-domain SAR image, are collected in 3D groups, subject to transform-domain shrinkage (Haar wavelet), reprojected in the image domain and aggregated.
Along the same lines, SAR-NN3D \cite{zeng2019computational} combines a pretrained CNN-based despeckler (CNN-SAR, ID-CNN) and nonlocal 3D shrinkage in the context of an iterative filtering procedure.
At the $k$-th iteration, the current denoised output is combined with the input and filtered again by the CNN, then similar blocks undergo nonlocal 3D shrinkage.
Unfortunately, no test on real SAR images is carried out.

Another iterative method is proposed in SAR-RDCP (recursive deep CNN prior) \cite{shen2020sar}, where filtering is cast as a MAP problem solved by means of half quadratic splitting.
The name of the method comes after the prior term,
estimated by means of a CNN which, to limit network complexity, uses the same parameters in all iterations, and is therefore subject to a recursive form of training.
The network itself is relatively small, employing dilated convolutions, residual blocks and channel attention modules.
The fidelity term, instead, is optimized by simple gradient descent.

In \cite{liu2019convolutional}, multiple denoisers, structurally similar to SAR-DRN, are trained in the log domain for various levels of noise.
Their outputs are then fused, using a saliency map computed on image details as an external guide.
The process is then iterated by filtering the resulting denoised image until a convergence criterion is met.
Extraordinarily good results are claimed, with 5-10 dB gains in PSNR over the state of the art, but the code is not available for replicating experiments.
A guide is used also in \cite{gu2020two}, where a small U-Net is adopted,
which takes as input not only the original image but also a texture level map (TLM) measuring the local homogeneity index as suggested in \cite{Gomez2016unassisted}.
The idea is that the texture level should support the denoising by providing a more accurate estimation of the local ENL.

Finally, in \cite{liu2020sar} and \cite{zhou2019deep}, multiresolution processing is considered.
In MSR-Net \cite{zhou2019deep}, three dyadic resolution levels are considered, and a multiresolution denoiser with CNN engine is used at each level.
The denoiser has a simple encoder-decoder architecture, but the parameters of the bottleneck layer flow to the upper resolution levels by means of a long short-term memory network, so as to enable fast convergence.
In \cite{liu2020sar}, instead, the input image is subject to a nonsubsampled shearlet transform with two levels of decomposition.
High-frequency coefficients are then denoised by means of a transform-domain method, while only low-frequency coefficients are denoised by using FFDNet in the log-domain.

\section{Supervised models: training and testing}
\label{sec:supervised_training}

Training is at the core of supervised deep learning methods.
The network learns to perform its task based on examples of the input paired with the desired output.
Such examples should be in large number, to ensure a good generalization and, needless to say, meaningful for the problem.
Unfortunately, no such thing as a clean SAR image exists, as already discussed in Section \ref{sec:data_avail}.
Therefore, approximate solutions are necessary to provide the network with an adequate number of reference images.

\subsection{Training procedures}

In principle, one could remove speckle, and hence obtain fully meaningful references, by means of temporal multilooking,
that is, by averaging a large number of co-registered images of the same scene, having the very same signal component but independent realizations of noise.
Indeed, this approach is followed in \cite{Chierchia2017SARCNN, cozzolino2019nonlocal, Cozzolino2020Nonlocal} where a stack of 26 single-look COSMO-SkyMed images is used for training with a leave-one-out strategy:
25 images are multilooked to provide the desired reference for the remaining one.
A similar procedure is used in \cite{tang2019sar} with a stack of 52 TerraSAR-X images.
This approach, though appealing, has two obvious limits.
First of all, 25 or even 50 images are not enough to adequately approximate a clean infinite-look reference.
A good despeckling filter can generate images that are very smooth in homogeneous areas, with an equivalent number of looks (ENL) easily exceeding 100.
Therefore, for these areas, a 25-look reference represents quite a poor example.
Not surprisingly, at a visual inspection, the images output by CNNs trained on multilook references are more effective at preserving high-frequency details than at suppressing speckle.
A second problem is that the signal must not change across the multitemporal stack.
This condition can be checked in advance, keeping for training only those regions that pass a suitable test.
Of course, the more images are used, the longer the temporal arc spanned, the less likely it is to find unchanged regions, so the two requirements are somewhat at odds with one another.

Another alternative is to use as clean references images despeckled by some other methods.
This approach does not really make sense if applied to individual images.
The trained network could, at best, mimic the original algorithm.
Moreover, current state-of-art methods cannot really provide filtered images of adequate quality for training, since speckle rejection is often obtained at the cost of resolution loss and filtering artifacts.
This procedure can be effective, instead, when used with data already filtered in the temporal domain.
Given a carefully multilooked image, one can apply a mild conservative despeckling filter to improve speckle rejection in homogeneous regions, without significant side effects.
This approach is indeed followed in \cite{yang2019sar, dalsasso2020sar},
where a relatively large stack of Sentinel-1 images is filtered in the temporal dimension, and the result is then filtered again in space applying MuLoG+BM3D.
A good quality reference is eventually obtained.
Contrary to previous methods, however, this clean reference is not paired with a noisy image of the stack for training, but with a simulated noisy image obtained by injecting speckle on the reference itself.
While this procedure provides many degrees of freedom for the experimental phase,
it creates poor noisy images, with speckle that is fully uncorrelated and affects also regions where the fully developed model does not hold.
Something similar is done in \cite{Lattari2019Deep}, where simulated speckle is injected on multilooked SAR data, without spatial filtering, to obtain noisy data for controlled experiments.

Apart from the above exceptions,
the vast majority of methods proposed thus far in the literature adopt a fully simulated training procedure.
Noiseless optical images are used as clean references, paired with simulated noisy images obtained through the injecting of white speckle.
The underlying assumption is that the trained models will eventually transfer well to the target domain.
However, the simulation procedure is flawed by several sources of inaccuracy, as already discussed in Section \ref{sec:data_avail}, and in fact experimental results provide only partial support to the use of fully simulated training.
In particular, the numerical performance observed on simulated test data does not seem to be a good predictor of despeckling quality in real SAR images.
Often, methods trained on synthetic data provide large improvements over the state-of-the-art on aligned data, but no apparent improvement on real-world SAR images.
Moreover, experiments carried out in \cite{dalsasso2020sar} confirm that, despite all the difficulties described above,
models trained on real SAR data guarantee a better despeckling quality than comparable models trained only on simulated data.

\subsection{Loss functions}

Let us now briefly analyze the loss functions proposed in the literature for the purpose of image despeckling.
Basically all methods include a standard data fidelity term, in most cases the $\ell_2$ distance (Euclidean distance, MSE, despeckling gain), but also the $\ell_1$ and smoothed-$\ell_1$ distance.
The relative merits of $\ell_1$ and $\ell_2$ losses are well known, with the former penalizing more harshly small errors and the latter focusing more on large errors.
There is no clear evidence, to date, in favor of one or the other for SAR despeckling.
Then, additional terms are sometimes included in the loss function.
A total variation loss is often used to ensure smooth solutions, sometimes leading to oversmoothing due to the relative simple prior (sparse image gradients) promoted by the total variation measure.
Methods relying on GANs, of course, include also an adversarial loss, based on a discriminator that decides if the despeckled image looks noiseless or not.
Of course, the discriminator (a CNN itself) will be only as good as the examples it sees in training, and can become a further source of inaccuracy in the absence of high quality references.
A perceptual loss term is used only in \cite{Wang2017generative},
with no ablation study in support, but it could be worth further analyses given the positive effects it produces in several computer vision problems.
In \cite{ferraioli2019novel, vitale2019new, vitale2020multi} the focus is almost exclusively on the loss function, with the aim to ensure better fidelity to the actual SAR statistics.
The network is regarded as an estimator of the image speckle content and the statistics of the removed speckle are compared with those of the ideal speckle.
Accordingly, a loss term is added which is the Kullback-Leibler divergence between the empirical and theoretical speckle pdf's.
Ablation studies show some improvements, especially for real SAR images.
This approach is lead to the extreme in \cite{yue2018sar} where, as mentioned in Section \ref{sec:supervised_architectures}, also the SAR signal is modeled as a random variable.
Accordingly, the loss function measures the Kullback-Leibler divergence between the pdf's of the observed and estimated SAR images (all in the log domain),
the latter computed as the convolution between the pdf's of the clean SAR image and speckle.

\subsection{Testing procedures}

We conclude this Section with a few words on testing procedures.
Almost always, the proposed methods are tested in two steps, first on simulated images, and then on real SAR images.

Considering that SAR image despeckling is the actual goal, the first step has limited significance, due to the poor transferability of models to the real target domain.
On the positive side, however, the presence of clean references allows for a sound assessment of performance,
based on widespread full-reference measures such as mean square error (MSE), peak signal-to-noise ratio (PSNR), structural similarity (SSIM).
Unfortunately, experiments are carried out on a wide variety of different test images, which prevents a direct comparison of numbers,
which is why we abstain from reporting any numerical results in this work.
An indirect (but somewhat shaky) comparison among methods could be established anyway based on results on common baselines with available code, like PPB and SAR-BM3D.
With a few notable exceptions, little or no ablation studies are carried out, experiments cover a limited range of cases (sometimes not even the single-look case) and the software is rarely published online.
In general,
there seems to be little attention to the reproducibility of research, with guaranteed negative impact on the speed of progress in this field.

All these problems only worsen when considering the real SAR data, given the absence of widespread datasets.
Nonetheless, they become irrelevant with respect to the absence of clean images, which prevents the use of full-reference measures.
Therefore, assessing the despeckling performance becomes a problem of its own, with many different solutions adopted in the literature.

The visual inspection of despeckled images is unanimously recognized as the foremost way of assessing image quality.
In fact, a despeckling algorithm should not only remove speckle, but also preserve image features (texture, edges, point targets, urban areas) and avoid introducing artifacts.
All these aspects are important, and not even a full reference measure could fully capture them all.
Notably, visual inspection is used in all papers reviewed here.
Unfortunately, being applied to different images of very different nature, possibly cherry-picked to underline some phenomena of interest, it becomes totally subjective, and useless for comparison purposes.
Also very popular, and useful, is the equivalent number of looks, computed as the local squared-mean to variance ratio.
Although it measures only speckle suppression, it does so in a very stable and reliable way, provided a large homogeneous region exists and is selected for measurements.

The {\it visual inspection of ratio images}, obtained as the pointwise ratio between original and despeckled images, is also precious.
Ideally, in homogeneous regions, the ratio image should contain only gamma-distributed weakly correlated speckle with no traces of the original image.
At a visual inspection, both signal leakages and noise correlation are easily detected in the ratio image, providing a subjective but reliable quality assessment tool.
Starting from the ratio image, one can also compute some simple numerical indexes,
like the Kullback-Leibler divergence between the empirical and theoretical distributions of speckle,
or the homogeneity index proposed in \cite{Gomez2016unassisted} to measure spatial correlation.

Many other ways to measure despeckling performance have been proposed in the literature, and their thorough analysis goes out of the scope of this paper.
We only mention briefly the benchmarking framework proposed in \cite{DiMartino2014benchmark} and available online.
Although relying itself on (physics-based) image simulation,
it tries to address explicitly the various aspects of despeckled image quality, from speckle rejection, to edge and texture preservation, to radiometric fidelity.
Towards this end, a set of canonical scenes is provided, both clean and noisy versions, together with a number of numerical measurement tools.
The weak point is that the selected scenes cannot account for the complexity of real SAR images.
However, the availability online of images and tools for objective assessment is the right way towards meaningful comparisons.

\section{Self-supervised models} \label{sec:selfsupervised}

The previous sections described the traditional avenue taken by research on deep learning despeckling algorithms, i.e., the supervised learning setting. As discussed, while the conceptual simplicity of the supervised setting is appealing, several issues stem from the lack of speckle-free ground truth data.
For this reason, the last year has witnessed a growing interest towards self-supervised methods, i.e., methods that can directly exploit noisy images without the need for clean data. This is an extremely important direction for the field as it allows to better exploit real SAR data, and do without optical-image datasets with their associated issues.  However, this new approach is still in its infancy and some of the works that will be presented are still undergoing peer review, so one must be careful with the information available at this stage. Nevertheless, the amount of publications and preprints \cite{ravani2019practical,Yuan2019BlindSI,yuan2019practical,Zhang2020Learning,dalsasso2020sar2sar,joo2019dopamine,Bordone2020IGARSS,Bordone2020speckle2void} on the topic that appeared in a short time signals a strong interest on these techniques and calls for further research. 

\begin{figure}[t]
\centering
\includegraphics[width=0.45\textwidth]{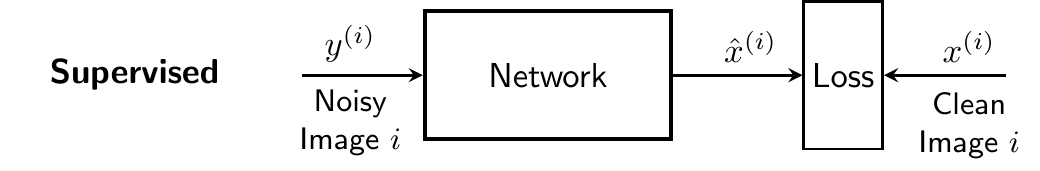}
\includegraphics[width=0.45\textwidth]{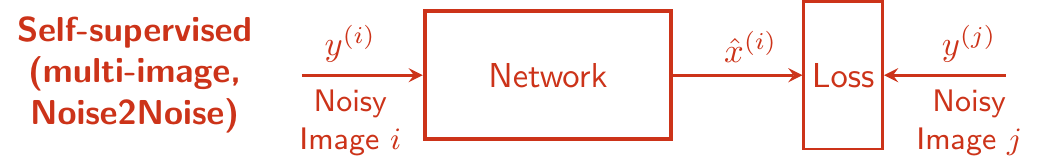}
\includegraphics[width=0.45\textwidth]{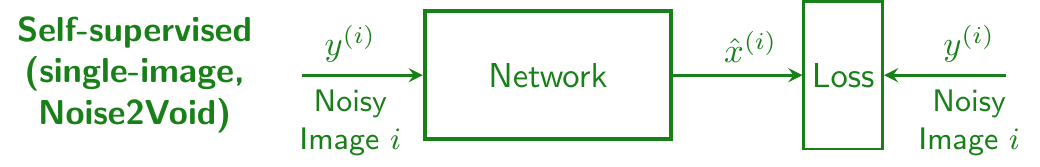}
\caption{Different approaches for model training. Supervised training requires a clean, speckle-free target. Self-supervised training uses noisy images as targets: either different realizations of the speckle process for the same scene (multi-image) or the same image (single-image).}
\label{fig:selfsupervised_block}
\end{figure}

Self-supervised models for SAR despeckling can be broadly categorized in two approaches, as shown in Fig. \ref{fig:selfsupervised_block}: multi-image models, and single-image models. These approaches are derived from recent techniques developed for traditional image denoising, which we now briefly review.

\subsection{Self-supervised models for image denoising}

Multi-image models follow the approach opened by Noise2Noise \cite{lehtinen2018noise2noise} whereby both the input and the training target are noisy images, representing the same scene with different realizations of the noise process. In standard supervised methods for image restoration, the model is trained by minimizing the distortion, usually measured using the $\ell_2$ norm, between the output and the ground truth image. In \cite{lehtinen2018noise2noise}, the authors observe that, on expectation, the estimate of the $\ell_2$ loss remains unchanged if we replace the ground truth images with noisy observations whose expectation matches the clean image. Consequently, as long as the expected value of the noisy images is equal to the ground truth image, it is possible to train a neural network using pairs of images with the same content but  independent realizations of noise, instead of clean-noisy image pairs. 
The need for multiple acquisitions of the same scene, however, represents a significant obstacle. 

\begin{figure}[t]
\centering
\includegraphics[width=0.3\textwidth]{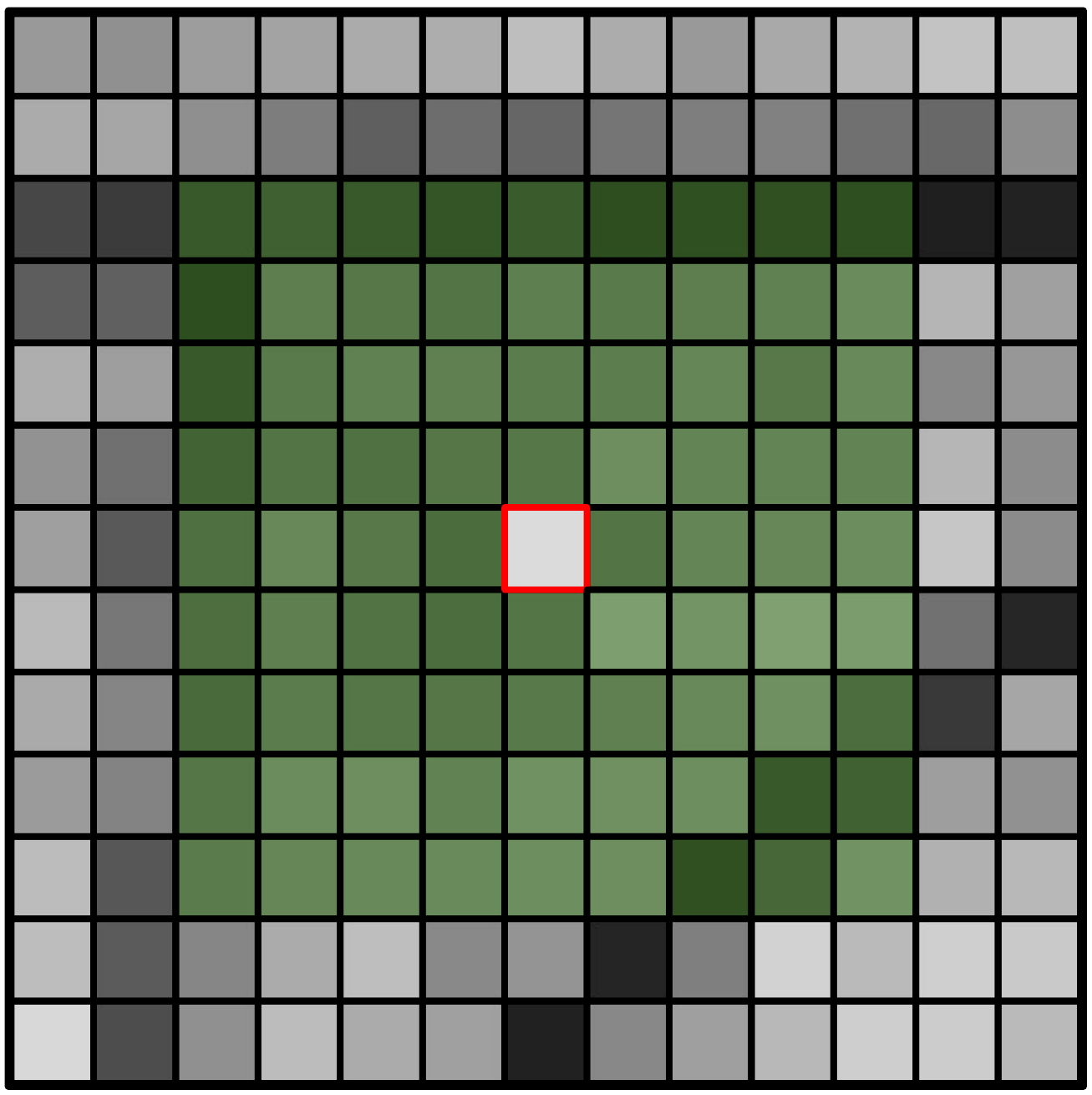}
\caption{Receptive field of a blind-spot network. The features associated to the pixels in green contribute to the features of the pixel with the red border. Notice how the pixel in red does not contribute to its own features.}
\label{fig:blindspot}
\end{figure}

This problem does not present itself in single-image techniques, which rely on careful data modeling, by assuming spatially uncorrelated noise and statistical priors about the data distribution. Two recent works, called Noise2Self \cite{batson2019noise2self} and Noise2Void \cite{krull2019noise2void}, show that, as long as the noise is spatially uncorrelated, it is possible to train an image restoration model using only individual noisy images. These two methods introduced the concept of blind-spot network, where a pixel is excluded from its own receptive field, as shown in Fig. \ref{fig:blindspot}. This allows the network to learn to estimate the center pixel from its receptive field, e.g., by minimizing the $\ell_2$ distance between the noisy pixel and its corresponding predicted value. By excluding the center pixel  from the receptive field, the network is prevented from learning the identity function. Clearly this works only when the noise is spatially uncorrelated, which may be challenging for SAR images.
Alternative single-image approaches also use losses that act as no-reference surrogates of the supervised loss (e.g., derived from SURE \cite{stein1981estimation}, which acts as an estimator of the MSE with respect to an unavailable clean image). Since all these single-image methods introduce a number of assumptions in order to avoid the need for multiple images, it remains to be seen how limiting these assumptions are and how they could be refined in future works.

\subsection{Self-supervised despeckling methods}
A few works tried to extend the Noise2Noise approach to despeckling. However, applying it in the despeckling context is not straightforward, since many acquisitions of the same scene are required. To overcome this issue different solutions have been proposed. In \cite{Yuan2019BlindSI} and \cite{yuan2019practical}, pairs of synthetically speckled images are generated. In particular, \cite{Yuan2019BlindSI} uses optical images with synthetic speckle, and so it is unclear why this method innovates with respect to supervised training, given that Noise2Noise comes with a performance penalty due to having the clean data only in the limit of infinitely many noisy realizations. Instead, \cite{yuan2019practical} is trained employing only real SAR images. In \cite{yuan2019practical}, an adversarial learning framework consisting in two generators and a discriminator is employed to produce images of the same scene with different speckle realizations. The adversarial training forces the distribution of the generated images to match that of the real SAR images. The main limitation of such type of solutions is that such matching is necessarily imperfect, leading to a domain gap between the generated images employed in training and the real images used in testing.  A different solution is presented in \cite{dalsasso2020sar2sar}, where multitemporal data is employed in order to obtain multiple images of the same scene. After a preliminary training with pairs of synthetically speckled images, the network is then fine-tuned  using a temporal series of SAR images, avoiding the domain gap problem that occurred in the previous methods. However, temporal changes might be present in the SAR time series and it is necessary to carefully compensate them. Authors' code is available online\footnote{\url{https://github.com/emanueledalsasso/SAR2SAR}}. Another work proposing a multitemporal approach is presented in \cite{ma2020sar}. In order to take into account temporal changes in the images of a time series, it is proposed to use a similarity measure for each input-reference pixel pair. Also in this case, authors' code is available online\footnote{\url{https://github.com/ahuyzx/NR-SAR-DL}}.

Even though the methods presented above propose different solutions to obtain multiple observations of the same scene, the problem still stands and represents the main limitation of the Noise2Noise approach. The works in \cite{Bordone2020IGARSS} and \cite{Bordone2020speckle2void} extend the single-image blind-spot approach of \cite{laine2019high} to propose a self-supervised Bayesian framework for SAR despeckling, including noise models and priors on the conditional distribution of the blind-spot pixel given its receptive field. The authors employ a whitening preprocessing and a network with variable blind-spot size in order to compensate for the autocorrelation of the speckle process. The main limitation of this approach lies in the assumptions that are necessarily introduced in order to train the model only with single noisy observations. Careful statistical modeling of the data and better handling of spatial correlation could unlock further improvements. Authors' code is available online\footnote{\url{https://github.com/diegovalsesia/speckle2void}}. 

Finally, DoPAMINE \cite{joo2019dopamine} also exploits a blind-spot network, not to directly compute the denoised pixel value, but to derive the parameters (slope and bias) of an affine regressor of each clean pixel given the noisy pixel. The loss function used to train the network is an unbiased estimator of the MSE derived from SURE, which only requires the noisy pixels and the despeckled output. As \cite{Bordone2020speckle2void}, the method heavily relies on spatially uncorrelated noise. Unfortunately, while the method could work with just real noisy images, the authors only present results by training on synthetically speckled images from the UC Merced Land use dataset.

\section{Future directions and open problems}
\label{sec:future_directions}

The analysis of existing work on SAR image despeckling based on deep learning has shown that the field has come a long way from its beginnings. However, we also feel that the available methods are not sufficiently mature yet, and that there is still a lot of work to be done towards methods that can provide despeckled images of consistently good quality. In the following we identify several aspects that should be subject of further work by the community. Some of these are related to the fact that works on the topic often do not adhere to the standards of reproducible research, making it difficult to assess the specific merits of each contribution with respect to the body of available literature, and significantly slowing down the pace of the technical advances.

\subsection{Software implementations}
		
Many papers claiming significant performance improvements over the state of the art do not include a freely available software implementation. Given the complexity of the most recent methods, as well as the effort to train a deep neural network in terms of time, computational resources and data, it is difficult for a researcher to implement several state-of-the-art methods from scratch in order to properly assess their own novel method. Authors should strive to make at least the trained models for their methods available, so that comparisons are possible. We note that, while this requires more effort on the authors' side, it typically leads to a greater impact of their work. Early dissemination, e.g. in the form of a preprint, also helps broadening the impact and speeding up the research pace, and in most venues it does not prevent from conference or journal publication.

\subsection{Common datasets} 

At the same time, even in case that deep models from existing papers are available, comparisons are only possible if the training and test datasets are also available. While it is indeed possible to re-train an existing network on a different dataset with respect to the one used by the original authors, this is not necessarily representative of the performance of the method. A fair comparison between two methods is only possible if they are trained and tested on the same data and using the same procedures. Therefore, for datasets that are publicly available, we encourage authors to clearly specify how the dataset has been used (e.g., training/validation/test splits). This also applies to the fully simulated approach, where it would be precious for the community to adopt common data and procedures for model training and testing.

Despite the inherent problems, and results that are still inconclusive,
we believe that using real SAR data is the main avenue to eventually generate good reference data.
Towards this end, one should exploit all available resources.
Multitemporal data are now freely available for some sensors, e.g., Sentinel-1, and collecting a large number of co-registered images should be relatively easy.
Moreover, optical remote sensing images co-registered with the SAR images are also available, e.g., Sentinel-2, and can be factored in through suitable fusion methods \cite{schmitt2018sen1}.
A careful combination of temporal multilooking, spatial despeckling and optical-SAR information fusion can certainly lead to high quality datasets for reliable training of deep-learning despeckling methods. For such derivative datasets, we encourage their owners to consider making them available to the community. 

\subsection{Common evaluation frameworks} 

We acknowledge that in some cases making software available may not be possible (e.g., due to company policy). Still, if researchers adopt common evaluation frameworks, it is possible for a researcher to compare the results obtained by their method with published results of other methods, if these results have been worked out in a standardized way on the same dataset and using the same quality metrics. An example is provided in \cite{DiMartino2014benchmark}; this paper provides a set of canonical scenes and corresponding simulated SAR images, along with corresponding objective measures on the SAR images that account for speckle suppression and feature preservation. This is an initial step towards standardizing performance evaluation and, while limited to a few specific cases, it indeed allows to provide a fair comparison among different methods. More in general, a first step towards building such common framework lies in the definition of a set of suitable quality metrics to be adopted, as the works in the available literature often employ different metrics.

\IEEEpeerreviewmaketitle



\begin{thebibliography}{100}
\providecommand{\url}[1]{#1}
\csname url@samestyle\endcsname
\providecommand{\newblock}{\relax}
\providecommand{\bibinfo}[2]{#2}
\providecommand{\BIBentrySTDinterwordspacing}{\spaceskip=0pt\relax}
\providecommand{\BIBentryALTinterwordstretchfactor}{4}
\providecommand{\BIBentryALTinterwordspacing}{\spaceskip=\fontdimen2\font plus
\BIBentryALTinterwordstretchfactor\fontdimen3\font minus
  \fontdimen4\font\relax}
\providecommand{\BIBforeignlanguage}[2]{{%
\expandafter\ifx\csname l@#1\endcsname\relax
\typeout{** WARNING: IEEEtran.bst: No hyphenation pattern has been}%
\typeout{** loaded for the language `#1'. Using the pattern for}%
\typeout{** the default language instead.}%
\else
\language=\csname l@#1\endcsname
\fi
#2}}
\providecommand{\BIBdecl}{\relax}
\BIBdecl

\bibitem{porcello1976}
L.~Porcello, N.~Massey, R.~Innes, and J.~Marks, ``Speckle reduction in
  synthetic-aperture radars,'' \emph{"J. Opt. Soc. Am."}, vol.~66, no.~11, pp.
  1305--1311, 1976.

\bibitem{kondo1977}
K.~Kondo, Y.~Ichioka, and T.~Suuki, ``Image restoration by {W}iener filtering
  in the presence of signal-dependent noise,'' \emph{"Applied Optics"},
  vol.~16, no.~9, pp. 2554--2558, 1977.

\bibitem{lee1980spatialadaptive}
J.-S. Lee, ``Digital image enhancement and noise filtering by use of local
  statistics,'' \emph{IEEE Transactions on Pattern Analysis and Machine
  Intelligence}, vol.~2, pp. 165--168, 1980.

\bibitem{kuan1985spatialadaptive}
D.~Kuan, A.~Sawchuk, T.~Strand, and P.~Chavel, ``Adaptive noise smoothing
  filter for images with signal-dependent noise,'' \emph{IEEE Transactions on
  Pattern Analysis and Machine Intelligence}, vol.~7, pp. 165--167, 1985.

\bibitem{frost1982spatialadaptive}
V.~Frost, J.~Stiles, K.~Shanmugan, and J.~Holtzman, ``A model for radar images
  and its application to adaptive digital filtering of multiplicative noise,''
  \emph{IEEE Transactions on Pattern Analysis and Machine Intelligence},
  vol.~4, pp. 157--166, 1982.

\bibitem{lopes1990spatialadaptive}
A.~Lopes, R.~Touzi, and E.~Nezry, ``Adaptive speckle filters and scene
  heterogeneity,'' \emph{IEEE Transactions on Geoscience and Remote Sensing},
  vol.~28, pp. 992--1000, 1990.

\bibitem{xie2002wavelet}
H.~Xie, L.~Pierce, and F.~Ulaby, ``{SAR} speckle reduction using wavelet
  denoising and {M}arkov random field modeling,'' \emph{IEEE Transactions on
  Geoscience and Remote Sensing}, vol.~40, pp. 2196--2212, 2002.

\bibitem{argenti2002wavelet}
F.~Argenti and L.~Alparone, ``Speckle removal from sar images in the
  undecimated wavelet domain,'' \emph{IEEE Transactions on Geoscience and
  Remote Sensing}, vol.~40, pp. 2363–--2374, 2002.

\bibitem{solbo2004wavelet}
S.~Solbo and T.~Eltoft, ``Homomorphic wavelet-based statistical despeckling of
  {SAR} images,'' \emph{IEEE Transactions on Geoscience and Remote Sensing},
  vol.~42, pp. 711--721, 2004.

\bibitem{bianchi2008wavelet}
T.~Bianchi, F.~Argenti, and L.~Alparone, ``Segmentation-based {MAP} despeckling
  of {SAR} images in the undecimated wavelet domain,'' \emph{IEEE Transactions
  on Geoscience and Remote Sensing}, vol.~46, pp. 2728–--2742, 2008.

\bibitem{aubert2008variational}
G.~Aubert and J.~Aujol, ``A variational approach to removing multiplicative
  noise,'' \emph{Siam Journal of Applied Mathematics}, vol.~68, pp. 925--946,
  2008.

\bibitem{bioucasdias2010variational}
J.~Bioucas-Dias and M.~Figueiredo, ``Multiplicative noise removal using
  variable splitting and constrained optimization,'' \emph{IEEE Transactions on
  Image Processing}, vol.~19, pp. 1720--1730, 2010.

\bibitem{shi2008variational}
J.~Shi and S.~Osher, ``A nonlinear inverse scale space method for a convex
  multiplicative noise model,'' \emph{Siam Journal of Imaging Science}, pp.
  294–--321, 2008.

\bibitem{deledalle2009nonlocal}
C.~Deledalle, L.~Denis, and F.~Tupin, ``Iterative weighted maximum likelihood
  denoising with probabilistic patch-based weights,'' \emph{IEEE Transactions
  on Image Processing}, vol.~18, pp. 2661–--2672, 2009.

\bibitem{Parrilli2012SARBM3D}
S.~Parrilli, M.~Poderico, C.~Angelino, and L.~Verdoliva, ``{A nonlocal SAR
  image denoising algorithm based on LLMMSE wavelet shrinkage},'' \emph{IEEE
  Transactions on Geoscience and Remote Sensing}, vol.~50, no.~2, pp. 606--616,
  Feb. 2012.

\bibitem{Cozzolino2014FANS}
D.~Cozzolino, S.~Parrilli, G.~Scarpa, G.~Poggi, and L.~Verdoliva, ``{Fast
  adaptive nonlocal SAR despeckling},'' \emph{IEEE Geoscience and Remote
  Sensing Letters}, vol.~11, no.~2, pp. 524--528, 2014.

\bibitem{deledalle2015nonlocal}
C.~Deledalle, L.~Denis, F.~Tupin, A.~Reigber, and M.~Jäger, ``{NL-SAR}: A
  unified nonlocal framework for resolution-preserving {(Pol)(In) SAR}
  denoising,'' \emph{IEEE Transactions on Geoscience and Remote Sensing},
  vol.~53, pp. 2021–--2038, 2015.

\bibitem{deledalle2014exploiting}
C.~Deledalle, L.~Denis, G.~Poggi, F.~Tupin, and L.~Verdoliva, ``Exploiting
  patch similarity for {SAR} image processing: the nonlocal paradigm,''
  \emph{IEEE Signal Processing Magazine}, vol.~31, pp. 69--78, 2014.

\bibitem{alexnet2012}
A.~Krizhevsky, I.~Sutskever, and G.~E. Hinton, ``Imagenet classification with
  deep convolutional neural networks,'' in \emph{Advances in Neural Information
  Processing Systems}, F.~Pereira, C.~J.~C. Burges, L.~Bottou, and K.~Q.
  Weinberger, Eds., vol.~25.\hskip 1em plus 0.5em minus 0.4em\relax Curran
  Associates, Inc., 2012, pp. 1097--1105.

\bibitem{zhang2017beyond}
K.~Zhang, W.~Zuo, Y.~Chen, D.~Meng, and L.~Zhang, ``Beyond a gaussian denoiser:
  Residual learning of deep cnn for image denoising,'' \emph{IEEE Transactions
  on Image Processing}, vol.~26, no.~7, pp. 3142--3155, 2017.

\bibitem{plotz2018neural}
T.~Pl{\"o}tz and S.~Roth, ``Neural nearest neighbors networks,'' in
  \emph{Advances in Neural Information Processing Systems}, 2018, pp.
  1087--1098.

\bibitem{liu2018non}
D.~Liu, B.~Wen, Y.~Fan, C.~C. Loy, and T.~S. Huang, ``Non-local recurrent
  network for image restoration,'' in \emph{Advances in Neural Information
  Processing Systems}, 2018, pp. 1673--1682.

\bibitem{valsesia2020deep}
D.~Valsesia, G.~Fracastoro, and E.~Magli, ``Deep graph-convolutional image
  denoising,'' \emph{IEEE Transactions on Image Processing}, vol.~29, pp.
  8226--8237, 2020.

\bibitem{dabov2007image}
K.~Dabov, A.~Foi, V.~Katkovnik, and K.~Egiazarian, ``Image denoising by sparse
  {3D} transform-domain collaborative filtering,'' \emph{IEEE Transactions on
  Image Processing}, vol.~16, no.~8, pp. 2080--2095, Aug. 2007.

\bibitem{Frery1997amodel}
A.~Frery, H.-J. Muller, C.~Yanasse, and S.~Sant’Anna, ``A model for extremely
  heterogeneous clutter,'' \emph{IEEE Transactions on Geoscience and Remote
  Sensing}, 1997.

\bibitem{pierce2002statistical}
H.~Xie, L.~Pierce, and F.~Ulaby, ``Statistical properties of logarithmically
  transformed speckle,'' \emph{IEEE Transactions on Geoscience and Remote
  Sensing}, 2002.

\bibitem{lapini2014blind}
A.~{Lapini}, T.~{Bianchi}, F.~{Argenti}, and L.~{Alparone}, ``{Blind speckle
  decorrelation for SAR image despeckling},'' \emph{IEEE Transactions on
  Geoscience and Remote Sensing}, vol.~52, no.~2, pp. 1044--1058, Feb 2014.

\bibitem{goodfellow2014generative}
I.~Goodfellow, J.~Pouget-Abadie, M.~Mirza, B.~Xu, D.~Warde-Farley, S.~Ozair,
  A.~Courville, and Y.~Bengio, ``Generative adversarial nets,'' in
  \emph{Advances in neural information processing systems}, 2014, pp.
  2672--2680.

\bibitem{kingma2014adam}
D.~P. Kingma and J.~Ba, ``Adam: A method for stochastic optimization,'' in
  \emph{International Conference on Learning Representations}, 2015.

\bibitem{Wang2017SAR}
P.~{Wang}, H.~{Zhang}, and V.~M. {Patel}, ``{SAR} image despeckling using a
  convolutional neural network,'' \emph{IEEE Signal Processing Letters},
  vol.~24, no.~12, pp. 1763--1767, Dec 2017.

\bibitem{Bordone2020speckle2void}
A.~{Bordone Molini}, D.~{Valsesia}, G.~{Fracastoro}, and E.~{Magli},
  ``{{Speckle2Void}: Deep Self-Supervised {SAR} Despeckling with Blind-Spot
  Convolutional Neural Networks},'' \emph{arXiv e-prints}, Jul. 2020.

\bibitem{mao2016image}
X.~Mao, C.~Shen, and Y.-B. Yang, ``Image restoration using very deep
  convolutional encoder-decoder networks with symmetric skip connections,'' in
  \emph{Advances in Neural Information Processing Systems}, 2016, pp.
  2802--2810.

\bibitem{tai2017memnet}
Y.~Tai, J.~Yang, X.~Liu, and C.~Xu, ``Memnet: A persistent memory network for
  image restoration,'' in \emph{IEEE International Conference on Computer
  Vision}, 2017, pp. 4539--4547.

\bibitem{MartinFTM01}
D.~Martin, C.~Fowlkes, D.~Tal, and J.~Malik, ``A database of human segmented
  natural images and its application to evaluating segmentation algorithms and
  measuring ecological statistics,'' in \emph{Proc. 8th Int'l Conf. Computer
  Vision}, vol.~2, July 2001, pp. 416--423.

\bibitem{gretton2009covariate}
A.~Gretton, A.~Smola, J.~Huang, M.~Schmittfull, K.~Borgwardt, and
  B.~Sch{\"o}lkopf, \emph{Covariate shift and local learning by distribution
  matching}.\hskip 1em plus 0.5em minus 0.4em\relax MIT Press, 2009, pp.
  131--160.

\bibitem{sentinel-1}
\BIBentryALTinterwordspacing
{Sentinel-1} data description. [Online]. Available:
  \url{https://sentinel.esa.int/web/sentinel/technical-guides/sentinel-1-sar/products-algorithms}
\BIBentrySTDinterwordspacing

\bibitem{terrasar-x}
\BIBentryALTinterwordspacing
{TerraSar-X} data description. [Online]. Available:
  \url{https://earth.esa.int/eogateway/catalog/terrasar-x-esa-archive}
\BIBentrySTDinterwordspacing

\bibitem{cosmoskymed}
\BIBentryALTinterwordspacing
{Cosmo-SkyMed} data description. [Online]. Available:
  \url{https://earth.esa.int/eogateway/catalog/cosmo-skymed-esa-archive}
\BIBentrySTDinterwordspacing

\bibitem{radarsat2}
\BIBentryALTinterwordspacing
{RADARSAT-2} data description. [Online]. Available:
  \url{https://earth.esa.int/eogateway/catalog/radarsat-2-esa-archive}
\BIBentrySTDinterwordspacing

\bibitem{airsar}
\BIBentryALTinterwordspacing
{AIRSAR} data description. [Online]. Available:
  \url{https://asf.alaska.edu/data-sets/sar-data-sets/airsar/}
\BIBentrySTDinterwordspacing

\bibitem{alos}
\BIBentryALTinterwordspacing
{ALOS-PALSAR} data description. [Online]. Available:
  \url{https://earth.esa.int/eogateway/catalog/alos-palsar-products}
\BIBentrySTDinterwordspacing

\bibitem{ers-2}
\BIBentryALTinterwordspacing
{AIRSAR} data description. [Online]. Available:
  \url{https://asf.alaska.edu/data-sets/sar-data-sets/ers-2/}
\BIBentrySTDinterwordspacing

\bibitem{isola2017image}
P.~Isola, J.-Y. Zhu, T.~Zhou, and A.~A. Efros, ``Image-to-image translation
  with conditional adversarial networks,'' in \emph{IEEE conference on computer
  vision and pattern recognition (CVPR)}, 2017, pp. 1125--1134.

\bibitem{yang2010bag}
Y.~Yang and S.~Newsam, ``Bag-of-visual-words and spatial extensions for
  land-use classification,'' in \emph{18th SIGSPATIAL International Conference
  on Advances in Geographic Information Systems}, 2010, pp. 270--279.

\bibitem{Cheng_2017}
G.~Cheng, J.~Han, and X.~Lu, ``Remote sensing image scene classification:
  Benchmark and state of the art,'' \emph{Proceedings of the IEEE}, vol. 105,
  no.~10, pp. 1865--1883, Oct 2017.

\bibitem{Chierchia2017SARCNN}
G.~Chierchia, D.~Cozzolino, G.~Poggi, and L.~Verdoliva, ``{SAR image
  despeckling through convolutional neural networks},'' in \emph{IEEE
  International Geoscience and Remote Sensing Symposium (IGARSS)}, 2017, pp.
  5438--5441.

\bibitem{yang2019sar}
X.~Yang, L.~Denis, F.~Tupin, and W.~Yang, ``{SAR} image despeckling using
  pre-trained convolutional neural network models,'' in \emph{Joint Urban
  Remote Sensing Event (JURSE)}, 2019, pp. 1--4.

\bibitem{dalsasso2020sar}
E.~Dalsasso, L.~Denis, and F.~Tupin, ``{SAR} image despeckling by deep neural
  networks: from a pre-trained model to an end-to-end training strategy,''
  \emph{arXiv preprint arXiv:2006.15559}, 2020.

\bibitem{pan2019filter}
T.~Pan, D.~Peng, W.~Yang, and H.-C. Li, ``A filter for {SAR} image despeckling
  using pre-trained convolutional neural network model,'' \emph{Remote
  Sensing}, vol.~11, no.~20, p. 2379, 2019.

\bibitem{yue2018sar}
D.-X. Yue, F.~Xu, and Y.-Q. Jin, ``{SAR} despeckling neural network with
  logarithmic convolutional product model,'' \emph{International journal of
  remote sensing}, vol.~39, no.~21, pp. 7483--7505, 2018.

\bibitem{Zhang2018learning}
Q.~Zhang, Q.~Yuan, J.~Li, Z.~Yang, X.~Ma, H.~Shen, and L.~Zhang, ``Learning a
  dilated residual network for {SAR} image despeckling,'' \emph{Remote
  Sensing}, vol.~10, pp. 1--18, february 2018.

\bibitem{Li2018HDRANet}
J.~Li, Y.~Li, Y.~Xiao, and Y.~Bai, ``{HDRANet}: Hybrid dilated residual
  attention network for {SAR} image despeckling,'' \emph{Remote Sensing},
  vol.~11, p. 2921, 12 2019.

\bibitem{Gui2018SAR}
Y.~Gui, L.~Xue, and X.~Li, ``{SAR} image despeckling using a dilated densely
  connected network,'' \emph{Remote Sensing Letters}, vol.~9, pp. 857--866,
  september 2018.

\bibitem{ZhangJing2019SIDU}
J.~Zhang, W.~Li, and Y.~Li, ``{SAR} image despeckling using multiconnection
  network incorporating wavelet features,'' \emph{IEEE Geoscience and Remote
  Sensing Letters}, pp. 1--5, 2019.

\bibitem{Lattari2019Deep}
F.~Lattari, B.~Leon, F.~Asaro, A.~Rucci, C.~Prati, and M.~Matteucci, ``Deep
  learning for {SAR} image despeckling,'' \emph{Remote Sensing}, vol.~11, p.
  1532, june 2019.

\bibitem{gu2017residual}
F.~Gu, H.~Zhang, C.~Wang, and B.~Zhang, ``Residual encoder-decoder network
  introduced for multisource {SAR} image despeckling,'' in \emph{SAR in Big
  Data Era: Models, Methods and Applications (BIGSARDATA)}, 2017, pp. 1--5.

\bibitem{zeng2018speckle}
T.~Zeng, Z.~Ren, and E.~Y. Lam, ``Speckle suppression using the convolutional
  neural network with an exponential linear unit,'' in \emph{Computational
  Optical Sensing and Imaging}, 2018, pp. CW5B--3.

\bibitem{bai2018fractional}
Y.-C. Bai, S.~Zhang, M.~Chen, Y.-F. Pu, and J.-L. Zhou, ``A fractional total
  variational {CNN} approach for {SAR} image despeckling,'' in
  \emph{International Conference on Intelligent Computing}, 2018, pp. 431--442.

\bibitem{Wang2017generative}
P.~{Wang}, H.~{Zhang}, and V.~M. {Patel}, ``Generative adversarial
  network-based restoration of speckled {SAR} images,'' in \emph{IEEE 7th
  International Workshop on Computational Advances in Multi-Sensor Adaptive
  Processing (CAMSAP)}, 2017, pp. 1--5.

\bibitem{liu2020gan}
R.~Liu, Y.~Li, and L.~Jiao, ``{SAR} image speckle reduction based on a
  generative adversarial network,'' in \emph{International Joint Conference on
  Neural Networks (IJCNN)}, 2020, pp. 1--6.

\bibitem{liu2020spatial}
Z.~Liu, R.~Lai, and J.~Guan, ``Spatial and transform domain cnn for sar image
  despeckling,'' \emph{IEEE Geoscience and Remote Sensing Letters}, 2020.

\bibitem{ferraioli2019novel}
G.~Ferraioli, V.~Pascazio, and S.~Vitale, ``A novel cost function for
  despeckling using convolutional neural networks,'' in \emph{Joint Urban
  Remote Sensing Event (JURSE)}, 2019, pp. 1--4.

\bibitem{vitale2019new}
S.~Vitale, G.~Ferraioli, and V.~Pascazio, ``A new ratio image based {CNN}
  algorithm for {SAR} despeckling,'' in \emph{IEEE International Geoscience and
  Remote Sensing Symposium (IGARSS)}, 2019, pp. 9494--9497.

\bibitem{vitale2020multi}
------, ``Multi-objective {CNN} based algorithm for {SAR} despeckling,''
  \emph{arXiv preprint arXiv:2006.09050}, 2020.

\bibitem{tang2019sar}
X.~Tang, L.~Zhang, and X.~Ding, ``{SAR} image despeckling with a multilayer
  perceptron neural network,'' \emph{International Journal of Digital Earth},
  vol.~12, no.~3, pp. 354--374, 2019.

\bibitem{zheng2018sar}
T.~Zheng, J.~Wang, and P.~Lei, ``{SAR} image despeckling based on variance
  constrained convolutional neural network,'' in \emph{Image and Signal
  Processing for Remote Sensing XXIV}, vol. 10789, 2018, p. 1078924.

\bibitem{cozzolino2019nonlocal}
D.~Cozzolino, L.~Verdoliva, G.~Scarpa, and G.~Poggi, ``Nonlocal {SAR} image
  despeckling by convolutional neural networks,'' in \emph{IEEE International
  Geoscience and Remote Sensing Symposium (IGARSS)}, 2019, pp. 5117--5120.

\bibitem{Cozzolino2020Nonlocal}
------, ``Nonlocal {CNN} {SAR} image despeckling,'' \emph{Remote Sensing},
  vol.~12, p. 1006, march 2020.

\bibitem{denis2019patches}
L.~Denis, C.-A. Deledalle, and F.~Tupin, ``From patches to deep learning:
  combining self-similarity and neural networks for {SAR} image despeckling,''
  in \emph{IGARSS 2019-2019 IEEE International Geoscience and Remote Sensing
  Symposium}, 2019, pp. 5113--5116.

\bibitem{zeng2019computational}
T.~Zeng, H.~K.-H. So, and E.~Y. Lam, ``Computational image speckle suppression
  using block matching and machine learning,'' \emph{Applied optics}, vol.~58,
  no.~7, pp. B39--B45, 2019.

\bibitem{shen2020sar}
H.~Shen, C.~Zhou, J.~Li, and Q.~Yuan, ``{SAR} image despeckling employing a
  recursive deep {CNN} prior,'' \emph{IEEE Transactions on Geoscience and
  Remote Sensing}, 2020.

\bibitem{liu2019convolutional}
S.~Liu, T.~Liu, L.~Gao, H.~Li, Q.~Hu, J.~Zhao, and C.~Wang, ``Convolutional
  neural network and guided filtering for {SAR} image denoising,'' \emph{Remote
  Sensing}, vol.~11, no.~6, p. 702, 2019.

\bibitem{gu2020two}
F.~Gu, H.~Zhang, and C.~Wang, ``A two-component deep learning network for {SAR}
  image denoising,'' \emph{IEEE Access}, vol.~8, pp. 17\,792--17\,803, 2020.

\bibitem{liu2020sar}
S.~Liu, L.~Gao, Y.~Lei, M.~Wang, Q.~Hu, X.~Ma, and Y.-D. Zhang, ``{SAR} speckle
  removal using hybrid frequency modulations,'' \emph{IEEE Transactions on
  Geoscience and Remote Sensing}, 2020.

\bibitem{zhou2019deep}
Y.~Zhou, J.~Shi, X.~Yang, C.~Wang, D.~Kumar, S.~Wei, and X.~Zhang, ``Deep
  multi-scale recurrent network for synthetic aperture radar images
  despeckling,'' \emph{Remote Sensing}, vol.~11, no.~21, p. 2462, 2019.

\bibitem{He2016residual}
K.~He, X.~Zhang, S.~Ren, and J.~Sun, ``Deep residual learning for image
  recognition,'' in \emph{IEEE conference on Computer Vision and Pattern
  Recognition}, 2016, pp. 770--778.

\bibitem{deledalle2017mulog}
C.~Deledalle, L.~Denis, S.~Tabti, and F.~Tupin, ``{MuLoG}, or how to apply
  {Gaussian} denoisers to multi-channel {SAR} speckle reduction?'' \emph{IEEE
  Transactions on Image Processing}, vol.~26, pp. 4389–--4403, 2017.

\bibitem{zhang2017ffdnet}
K.~Zhang, W.~Zuo, and L.~Zhang, ``{FFDNet: Toward a fast and flexible solution
  for CNN-based image denoising},'' \emph{IEEE Transactions on Image
  Processing}, vol.~27, pp. 4608--4622, 2017.

\bibitem{huang2017densenet}
G.~Huang, Z.~Liu, L.~Van Der~Maaten, and K.~Q. Weinberger, ``Densely connected
  convolutional networks,'' in \emph{IEEE conference on computer vision and
  pattern recognition (CVPR)}, 2017, pp. 2261--2269.

\bibitem{ronneberger2015unet}
O.~Ronneberger, P.~Fischer, and T.~Brox, ``U-net: Convolutional networks for
  biomedical image segmentation,'' in \emph{International Conference on Medical
  Image Computing and Computer-Assisted Intervention (MICCAI)}, 2015, pp.
  234--241.

\bibitem{johnson2015perceptual}
J.~Johnson, A.~Alahi, and F.-F. L., ``Perceptual losses for realtime style
  transfer and super-resolution,'' in \emph{European Conference on Computer
  Vision (ECCV)}, 2016.

\bibitem{simonyan2015verydeep}
K.~Simonyan and A.~Zisserman, ``Very deep convolutional networks for
  large-scale image recognition,'' in \emph{International Conference on
  Learning Representations}, 2015.

\bibitem{deledalle2009iterative}
C.~{Deledalle}, L.~{Denis}, and F.~{Tupin}, ``Iterative weighted maximum
  likelihood denoising with probabilistic patch-based weights,'' \emph{IEEE
  Transactions on Image Processing}, vol.~18, no.~12, pp. 2661--2672, Dec 2009.

\bibitem{lefkimmiatis2017nonlocal}
S.~Lefkimmiatis, ``Non-local color image denoising with convolutional neural
  networks,'' in \emph{IEEE Conference on Computer Vision and Pattern
  Recognition (CVPR)}, 2017, pp. 5882–--5891.

\bibitem{wang2018nonlocal}
X.~Wang, R.~Girshick, A.~Gupta, and K.~He, ``Non-local neural networks,'' in
  \emph{IEEE Conference on Computer Vision and Pattern Recognition (CVPR)},
  2018.

\bibitem{cruz2018nonlocality}
C.~Cruz, A.~Foi, V.~Katkovnik, and K.~Egiazarian, ``Nonlocality-reinforced
  convolutional neural networks for image denoising,'' \emph{IEEE Signal
  Processing Letters}, vol.~25, no.~8, pp. 1216--1220, 2018.

\bibitem{Gomez2016unassisted}
L.~Gomez, R.~Ospina, and F.~A., ``Unassisted quantitative evaluation of
  despeckling filters,'' \emph{Remote Sensing}, vol.~9, no.~4, April 2017.

\bibitem{DiMartino2014benchmark}
G.~D. Martino, M.~Poderico, G.~Poggi, D.~Riccio, and L.~Verdoliva,
  ``{Benchmarking framework for SAR despeckling},'' \emph{IEEE Transactions on
  Geoscience and Remote Sensing}, vol.~52, no.~3, pp. 1596--1615, 2014.

\bibitem{ravani2019practical}
K.~Ravani, S.~Saboo, and J.~S. Bhatt, ``A practical approach for {SAR} image
  despeckling using deep learning,'' in \emph{IEEE International Geoscience and
  Remote Sensing Symposium (IGARSS)}, 2019, pp. 2957--2960.

\bibitem{Yuan2019BlindSI}
Y.~Yuan, J.~Sun, and J.~Guan, ``{Blind {SAR} Image Despeckling Using
  Self-Supervised Dense Dilated Convolutional Neural Network},'' \emph{ArXiv},
  vol. abs/1908.01608, 2019.

\bibitem{yuan2019practical}
Y.~Yuan, J.~Sun, J.~Guan, P.~Feng, and Y.~Wu, ``A practical solution for {SAR}
  despeckling with only single speckled images,'' \emph{arXiv preprint
  arXiv:1912.06295}, 2019.

\bibitem{Zhang2020Learning}
G.~Zhang, Z.~Li, X.~Li, and Y.~Xu, ``{Learning synthetic aperture radar image
  despeckling without clean data},'' \emph{Journal of Applied Remote Sensing},
  vol.~14, no.~2, pp. 1 -- 20, 2020.

\bibitem{dalsasso2020sar2sar}
E.~Dalsasso, L.~Denis, and F.~Tupin, ``{SAR2SAR}: a self-supervised despeckling
  algorithm for {SAR} images,'' \emph{arXiv preprint arXiv:2006.15037}, 2020.

\bibitem{joo2019dopamine}
S.~Joo, S.~Cha, and T.~Moon, ``{DoPAMINE}: Double-sided masked {CNN} for pixel
  adaptive multiplicative noise despeckling,'' in \emph{AAAI Conference on
  Artificial Intelligence}, vol.~33, 2019, pp. 4031--4038.

\bibitem{Bordone2020IGARSS}
A.~{Bordone Molini}, D.~{Valsesia}, G.~{Fracastoro}, and E.~{Magli}, ``{Towards
  Deep Unsupervised {SAR} Despeckling with Blind-Spot Convolutional Neural
  Networks},'' in \emph{IEEE International Geoscience and Remote Sensing
  Symposium (IGARSS)}, 2020.

\bibitem{lehtinen2018noise2noise}
J.~Lehtinen, J.~Munkberg, J.~Hasselgren, S.~Laine, T.~Karras, M.~Aittala, and
  T.~Aila, ``Noise2noise: Learning image restoration without clean data,'' in
  \emph{International Conference on Machine Learning}, 2018, pp. 2965--2974.

\bibitem{batson2019noise2self}
J.~Batson and L.~Royer, ``Noise2self: Blind denoising by self-supervision,'' in
  \emph{International Conference on Machine Learning}, 2019, pp. 524--533.

\bibitem{krull2019noise2void}
A.~Krull, T.-O. Buchholz, and F.~Jug, ``Noise2void-learning denoising from
  single noisy images,'' in \emph{IEEE Conference on Computer Vision and
  Pattern Recognition (CVPR)}, 2019, pp. 2129--2137.

\bibitem{stein1981estimation}
C.~M. Stein, ``Estimation of the mean of a multivariate normal distribution,''
  \emph{The annals of Statistics}, pp. 1135--1151, 1981.

\bibitem{ma2020sar}
X.~Ma, C.~Wang, Z.~Yin, and P.~Wu, ``Sar image despeckling by noisy
  reference-based deep learning method,'' \emph{IEEE Transactions on Geoscience
  and Remote Sensing}, 2020.

\bibitem{laine2019high}
S.~Laine, T.~Karras, J.~Lehtinen, and T.~Aila, ``High-quality self-supervised
  deep image denoising,'' in \emph{Advances in Neural Information Processing
  Systems}, 2019, pp. 6970--6980.

\bibitem{schmitt2018sen1}
M.~Schmitt, L.~Hughes, and X.~Zhu, ``The sen1-2 dataset for deep learning in
  sar-optical data fusion,'' \emph{ISPRS Annals of Photogrammetry, Remote
  Sensing \& Spatial Information Sciences}, vol.~4, no.~1, 2018.

\end{thebibliography}
\end{document}